\documentclass[10pt]{article}
\usepackage{amsmath}
\usepackage{graphicx}
\usepackage{amsfonts}
\usepackage{amssymb}
\usepackage{epsf}
\usepackage{latexsym}

\def\uRR{{\underline{R}}}

\def\uPP{{\underline{P}}}

\def\uiota{{\underline{\iota}}}
\def\80{\hspace{0.8in}}

\def\n{\mbox{n}}
\def\nm{\mbox{m}}

\def\nC{\mbox{C}}
\def\bm{\bf{m}}
\newcommand{\be}{\begin{enumerate}}
\newcommand{\ee}{\end{enumerate}}
\newcommand{\bi}{\begin{itemize}}
\newcommand{\ei}{\end{itemize}}
\newcommand{\bd}{\begin{description}}
\newcommand{\ed}{\end{description}}

\def\beq{\begin{equation}}
\def\eeq{\end{equation}}
\def\bea{\begin{eqnarray}}
\def\eea{\end{eqnarray}}
\def\hat{\widehat}

\def\half{\frac{1}{2}}
\def\pa{\partial}
\def\d{\textrm{d}}

\def\ttL{\mbox{\tt L}}
\def\ttD{\mbox{\tt D}}

\def\ttP{\mbox{\tt P}}
\def\ttQ{\mbox{\tt Q}}

\def\uttL{\underline{\mbox{\tt L}}}

\def\Circ{\mbox{\Large$\circ$}}
\def\Star{\mbox{\Large $\ast$}}

\def\sumin{\sum\mbox{}_{\mbox{}_{\mbox{\scriptsize $i$=1}}}^{\sn}}
\def\cr{\mbox{\scriptsize{\bf $\mbox{ } \times \mbox{ }$}}}

\def\bigr{\mbox{\Large\sffamily r}}

\def\nl{\mbox{l}}

\def\mn{\mbox{n}}

\def\nm{\mbox{m}}

\def\mA{\mbox{A}}
\def\mB{\mbox{B}} 
\def\mC{\mbox{C}}

\def\mF{\mbox{F}}

\def\mI{\mbox{I}}

\def\mN{\mbox{N}}

\def\sa{\mbox{\scriptsize a}}
\def\sb{\mbox{\scriptsize b}}
\def\sc{\mbox{\scriptsize c}}
\def\sd{\mbox{\scriptsize d}}
\def\se{\mbox{\scriptsize e}}
\def\sf{\mbox{\scriptsize f}}
\def\sg{\mbox{\scriptsize g}} 
\def\sh{\mbox{\scriptsize h}} 
\def\si{\mbox{\scriptsize i}}
 
\def\sk{\mbox{\scriptsize k}}
\def\sll{\mbox{\scriptsize l}} 
\def\sm{\mbox{\scriptsize m}}
\def\sn{\mbox{\scriptsize n}} 
\def\so{\mbox{\scriptsize o}} 
\def\sp{\mbox{\scriptsize p}}

\def\sr{\mbox{\scriptsize r}}

\def\sss{\mbox{\scriptsize s}}
\def\st{\mbox{\scriptsize t}}
\def\su{\mbox{\scriptsize u}}

\def\sx{\mbox{\scriptsize x}}

\def\sA{\mbox{\scriptsize A}} 
\def\sB{\mbox{\scriptsize B}}
\def\sC{\mbox{\scriptsize C}}
\def\sD{\mbox{\scriptsize D}}
\def\sE{\mbox{\scriptsize E}}
\def\sF{\mbox{\scriptsize F}}
\def\sG{\mbox{\scriptsize G}}
\def\sH{\mbox{\scriptsize H}}

\def\sK{\mbox{\scriptsize K}}
 
\def\sM{\mbox{\scriptsize M}} 
\def\sN{\mbox{\scriptsize N}} 

\def\sP{\mbox{\scriptsize P}} 
 
\def\sR{\mbox{\scriptsize R}}
\def\sS{\mbox{\scriptsize S}}
\def\sT{\mbox{\scriptsize T}}

\def\sW{\mbox{\scriptsize W}}
 
\def\sY{\mbox{\scriptsize Y}}

\def\barr{\bar{r}}

\def\eph(B){\mbox{\scriptsize emergent(LMB)}}

\def\ta{\mbox{\tiny a}}

\def\te{\mbox{\tiny e}}

\def\th{\mbox{\tiny h}}

\def\tl{\mbox{\tiny l}}
\def\tm{\mbox{\tiny m}}

\def\to{\mbox{\tiny o}}
\def\tp{\mbox{\tiny p}}

\def\ts{\mbox{\tiny s}}
\def\ttt{\mbox{\tiny t}}

\def\tF{\mbox{\tiny F}}

\def\tN{\mbox{\tiny N}}

\def\tS{\mbox{\tiny S}}
\def\tT{\mbox{\tiny T}}

\def\sbg{\mbox{{\bf \scriptsize g}}}

\def\bg{\mbox{{\bf g}}}
\def\bz{\mbox{{\bf z}}}

\def\tbg{\mbox{{\bf \tiny g}}}
\def\fE{\mbox{\sffamily E}}

\def\fH{\mbox{\sffamily H}}
\def\fI{\mbox{\sffamily I}}

\def\fK{\mbox{\sffamily K}}

\def\fP{\mbox{\sffamily P}}
\def\fQ{\mbox{\sffamily Q}}
\def\fR{\mbox{\sffamily R}}
\def\fS{\mbox{\sffamily S}}
\def\fT{\mbox{\sffamily T}}

\def\fW{\mbox{\sffamily W}}

\begin{document}
\begin{titlepage}
\vspace{.7in}
\begin{center}
 
\vspace{2in} 

{\LARGE\bf RELATIONAL MECHANICS OF SHAPE AND SCALE}

\vspace{.2in}

{\bf Edward Anderson}$^{1}$ 

\vspace{.2in}

{\em $^1$ DAMTP Cambridge U.K.}

\vspace{.2in}

\end{center}

\begin{abstract}

Relational particle mechanics models (RPM's) are useful models for the problem of time in quantum 
gravity and other foundational issues in quantum cosmology.   
Some concrete examples of scalefree RPM's have already been studied, but it is the case with scale that 
is needed for the semiclassical and dilational internal time approaches to the problem of time.   
In this paper, I show that the scaled RPM's configuration spaces are the cones over the scalefree RPM's 
configuration spaces, which are spheres in 1-$d$ and complex projective spaces in 2-$d$ for plain 
shapes, and these quotiented by $\mathbb{Z}_2$ for oriented shapes.  
I extend the method of physical interpretation by tessellation of the configuration space and the 
description in terms of geometrical quantities to the cases with scale and/or orientation.
I show that there is an absence of monopole issues for RPM's and point out a difference between 
quantum cosmological operator ordering and that used in molecular physics. 
I use up RPM's freedom of the form of the potential to more closely parallel various well-known 
cosmologies, and begin the investigation of the semiclassical approach to the problem of time for 
such models.

\end{abstract}

\vspace{0.2in}


\vspace{0.2in}


\mbox{ } 

%
%

\mbox{ } 


\mbox{ }

\noindent$^1$ ea212@cam.ac.uk \mbox{ } 

\mbox{ }

\end{titlepage}

\section{Introduction}

This paper concerns Leibniz--Mach--Barbour-type \cite{LMach, BB82, B94I, EOT} 
relationalism,\footnote{This 
is distinct from Rovelli's use of the word `relational' \cite{Rovelli}; see also \cite{08I} for a 
comparison between the two.}
which involves the following postulates and implementations.  

\noindent 
i) {\it Temporal relationalism} is that there is no overall notion of time for the universe as a whole. 
This is mathematically implemented at the classical level by considering reparametrization-invariant 
actions that are free of extraneous time variables [e.g. Newtonian time or General Relativity (GR)'s 
lapse].  
Such actions are often Jacobi-type actions \cite{Lanczos}, whose integrand is the square root of the 
product of kinetic and potential factors.
Reparametrization invariance then guarantees that primary constraints exist \cite{Dirac}, which for 
Jacobi-type actions have quadratic but not linear dependence in the momenta.  

\noindent ii) {\it Configurational relationalism} is that there is a certain group $G$ of 
transformations that are physically irrelevant, between the physically indiscernible states 
\cite{BB82, B03}.  
This can be implemented by considering one's action to be built out of arbitrary $G$-frame objects.  
In the principal examples of this paper, the auxiliary variables associated with $G$ then only show up 
explicitly in the action as corrections to its velocities, and variation with respect to these 
auxiliaries produces linear secondary constraints that subsequently ensure the physical irrelevance of 
$G$.

Relational particle mechanics (RPM's) are examples of relational theories in this sense.  
In {\it Euclidean RPM (ERPM)} \cite{BB82} (and further studied in \cite{ERPM, B94I, EOT, Paris, 06I, 
TriCl, 08I}) $G$ is the Euclidean group of translations and rotations, so that this is a mechanics in 
which only relative times, relative angles and relative separations are meaningful.  
On the other hand, in {\it similarity RPM (SRPM)} (proposed in \cite{B03} and on distinct foundations in 
\cite{FORD} and further studied in \cite{SRPM, Paris, 06II, TriCl, 08I, 08II, AF, +tri}), $G$ is the 
similarity group of translations, rotations and dilations, so that this is a mechanics in which only 
relative times, relative angles and ratios of relative separations are meaningful.

The more recently proposed SRPM turns out to be easier to study.  
For, what turned out to be \cite{FORD} a configuration space study for SRPM was found to be already 
present elsewhere in the literature \cite{Kendall}.  
Furthermore in 1- and 2-$d$ (I term these models with N particles {\it N-stop metroland} and 
{\it N-a-gonland} respectively, referring to the 3 and 4 particle cases of the latter as  
{\it triangleland} and {\it quadrilateralland}), the SRPM configuration spaces are 
$\mathbb{S}^{\sN - 2}$ and $\mathbb{CP}^{\sN - 2}$.  
This gives a second direct way of implementing configurational relationalism as the natural mechanics in 
the sense of Jacobi and of Synge (see e.g. \cite{Lanczos}) on these geometries. 
And, via numerous geometrical and methods of mathematical physics techniques being available, it also 
permits solution of concrete examples \cite{08I, 08II, MGM, AF, +tri}.  
However, the below motivation makes it clear that it is ERPM that we want for many purposes; the 
present paper concerns setting up the study of this.  
The abovementioned SRPM study continues to be of value here due to its also occuring as a subproblem 
within ERPM.   
My minor motivation for studying RPM's concerns the absolute versus relative motion debate 
that goes back to Newton and Leibniz \cite{LMach, bucket} (while this is 
very important topic, we only have a few contributions to make to it in this paper).  
My major motivation, however, is as follows.
GR is also relational in the above sense, since it can be cast in terms of a Jacobi-type 
action with the spatial 3-diffeomorphisms playing the role of $G$ \cite{BSW, RWR, OMSan, ABFKO, Paris, 
06II, FORD, FEPI, 08II, Banal} -- this is a variant of its canonical geometrodynamics formulation 
\cite{ADM, Dirac, DeWitt67, Battelle}.  
This furbishes the large number of analogies in Secs 4 and 7, by which RPM's are a useful toy model of 
(the canonical geometrodynamics formulation of) GR in many ways.   
For example RPM's have recently been studied as regards building fairly solvable models of quantization 
approaches \cite{SRGryb, 06I, 08II, Banal}. 
%
%
RPM's relational features are useful for whole-universe modelling, which is the setting for Quantum 
Cosmology and the Problem of Time in Quantum Gravity.  
For both of these motivations, then, working with ERPM rather than SRPM is preferable due to scale  
usually being taken to be physically significant in nature.

The notorious {\it Problem of Time} \cite{K81, K91, K92, I93, K99, Kiefer, Smolin08} occurs because 
`time' takes a different meaning in each of GR and ordinary Quantum Theory.  
This incompatibility underscores a number of problems with trying to replace these two branches with a 
single framework in situations in which the premises of both apply, namely in black holes and in the 
very early universe.  
One facet of the Problem of Time appears in attempting canonical quantization of GR in its dynamical 
guise of `geometrodynamics' (evolving spatial geometries).
This gives a GR momentum constraint ${\cal L}_{\mu}$ that is linear in the momenta as well as a 
Hamiltonian constraint ${\cal H}$ that is quadratic but not linear in the momenta. 
Then elevating ${\cal H}$ to a quantum equation produces a stationary i.e timeless or frozen wave 
equation --- the  Wheeler-DeWitt equation $\hat{\cal H}\Psi = 0$ (for $\Psi$ the wavefunction of the 
Universe) --- in place of ordinary QM's time-dependent one.
See Sec 5 (or \cite{K92, I93} in more detail) for other facets of the Problem of Time.

There are various distinct and conceptually-interesting strategies proposed toward understanding the 
Problem of Time. 
a) Perhaps one is to seek a {\it hidden time} within classical GR by canonical transformation 
\cite{York72Time, K81, K92, I93}. 
b) Or perhaps GR in general has no time fundamentally, but a time nevertheless {\it emerges} under 
certain circumstances, e.g. in the {\it semiclassical approach}'s regime, in which slow, heavy `H' 
variables provide an approximate timestandard with respect to which the other fast, light `L' degrees of 
freedom evolve \cite{HallHaw,K92, Kiefer}. 
In GR Quantum Cosmology, `H' is scale (and homogeneous matter modes) and `L' are inhomogeneities. 
c) Or perhaps one should take the universe as a whole to be timeless \cite{PW83, GMH, EOT, H99, Records} 
and see what can be done, e.g. considering only questions about the universe `being', rather than 
`becoming', a certain way.  
E.g. {\it records theory} \cite{GMH, H99, B94II, EOT, Records} concerns whether 
localized subconfigurations of a single instant contain useable information/correlations, and whether a
semblance of dynamics or history thereby arises. 
d) Or, perhaps instead it is the histories that are primary: {\it histories theory} \cite{GMH, Hartle}.  
None of the above strategies has been carried out detail for full GR; they are usually probed with toy 
models.

Minisuperspace \cite{Mini, Magic, HH83} (homogeneous GR) has often been used thus \cite{K92, I93}. 
This paper's RPM analogies include various that are rendered trivial in minisuperspace.  
An important feature of GR (and one missed out by minisuperspace models) is ${\cal L}_{\mu}$, and this 
causes substantial complications e.g. in attempted resolutions of the Problem of Time \cite{K92, I93}.  
However, RPM's zero total angular momentum constraint ${\ttL}_{\mu}$ (arising from variation with respect 
to the rotational auxiliary) is a nontrivial analogue of ${\cal L}_{\mu}$ in a number of ways. 
Another feature of GR that has an analogue for RPM's but not for minisuperspace is the possession of a 
notion of localization and thus of structure formation. 
Minisuperspace is, however, closer to GR I) in having more specific and GR-inherited potentials to RPM's 
much greater freedom in these: in SRPM the potential must be homogeneous of degree 0 (i.e. 
a function of pure shape alone), while in ERPM the potential is completely free.  
II) In having indefinite kinetic terms to RPM's mechanical and hence positive-definite ones. 
Thus minisuperspace and RPM's are to some extent complementary in their similarities to GR, and thus in 
their usefulnesses as toy models thereof. 
Midisuperspace has all of these features but at the price of considerable technical complexity, which 
obstructs a number of Problem of Time calculations.

The other foundational Quantum Cosmology issues that RPM's do or are likely to contribute to (at 
least qualitatively) are as follows.  
Does structure formation in the universe have a quantum mechanical origin \cite{HallHaw}?  
In GR, this requires midisuperspace or at least inhomogeneous perturbations about minisuperspace, 
and these are hard to study.   
There are also a number of difficulties associated with closed system physics \cite{DeWitt67, 06I} and 
observables, speculations on initial conditions, the meaning and form of the wavefunction of the 
universe (e.g. whether a uniform state is to play an important role \cite{Penrose, AF, +tri}) and 
attempting to explain the arrow of time \cite{HH83, EOT, H03, Rovelli}.

See \cite{BS89, SemiclI, 08II, MGM, AF, +tri, SemiclIII} for quantum cosmological use of RPM's and 
\cite{BS89, K92, I93, Kiefer, Paris, 06II, SemiclI, Records, 08II, BF08, AF, +tri, MGM, SemiclIII, 
scaleQM, 08III, NOD, NOI, IT} for uses of RPM's in the study of the Problem of Time.    


So far in the RPM toy model program, and mostly still so in the present paper, it is plain rather than 
oriented shapes that are considered (i.e. a shape and its mirror image are considered distinct).  
The main new feature of the present paper is inclusion of scale in the shape-scale split form of the (already) 
reduced formulation, so as to be able to start making contact with the aforementioned quantum 
cosmological issues, and semiclassical and dilational hidden time approaches to the Problem of Time.  
Scalefree 4-stop metroland \cite{AF} and scalefree triangleland \cite{08II, +tri} have already been 
studied.  
In the context with scale, 3-stop metroland also becomes meaningful (it now possesses two degrees of 
freedom so that it can be solved for one of these in terms of the other, time itself being meaningless 
in relational whole-universe models); the present paper studies this, scaled 4-stop metroland and 
scaled triangleland.  
Each of 4-stop metroland and triangleland differ in some useful features as regards Problem of Time 
and Quantum Cosmology applications.  
The advantage of triangleland is that it permits modelling of situations with both scale and a 
nontrivial constraint. 
The advantages of 4-stop metroland are an easier realization of the sphere, and that it can be split 
into subsystems, each of which is nontrivial (useful for records theory).
Quadrilateralland combines all of these advantages but is harder and beyond the scope of this paper 
(see \cite{QShape} for a start); it also makes good sense to study these advantages and their 
application to Problem of Time strategies separately in such regimes with more tractable (spherical) 
mathematics prior to attempting to combine them in an arena with rather less tractable 
($\mathbb{CP}^2$) mathematics.
A long term goal of the RPM program is to use quadrilateralland to investigate how to combine records theory with 
scale-driven semiclassical approach to the Problem of Time and to Quantum Cosmology (perhaps 
simultaneously with histories theory).
The interest in such a combination (see e.g. \cite{H03, GMH, H99}) is due to there being a records 
theory within histories theory, histories decohereing (`self-measuring') being one possible way of 
obtaining a semiclassical regime in the first place, and the elusive question of what decoheres what 
should be answerable through where information is actually stored, i.e. what and where the records 
thereby formed are.  


In scalefree RPM's, understanding the configuration spaces ({\it shape spaces}) \cite{FORD} was key to 
(and provided an alternative foundation for) the study \cite{08I, 08II, AF, +tri} of their Classical 
and Quantum Mechanics.    
Thus I likewise begin the present paper on scaled RPM (Sec 2) by considering its configuration spaces 
({\it relational spaces}).  
In shape--scale split form, these are the {\it cones} over the corresponding shape spaces.   
Appropriate mathematical treatment of shape space already existed in the literature in a 
different context (work by Kendall \cite{Kendall84, Kendall} toward the statistical theory of shape); 
the present paper gathers the scaled relational space counterpart with a number of ties to 
the Celestial Mechanics and Molecular Physics literatures.    
Following \cite{AF, +tri}, I furthermore consider how the shape spaces and relational spaces 
(which I jointy term {\it relationalspaces}) can be tessellated by their physical interpretation.  
This is useful for subsequent reading off the physical significance of classical trajectories and QM 
wavefuctions \cite{AF,+tri,scaleQM,08III,SemiclIII,Tpaper}.  
I also explain how to consider scaled 3- and 4-stop metroland and triangleland in terms of periodic 
quantities $u^{\Delta}$ whose squares sum to 1 and a scale variable running from 0 to $\infty$.
While the last two are both $\mathbb{S}^2$ and involve three $u^{\Delta}$'s, they differ significantly 
in the form and geometrical meaning that their $u^{\Delta}$ and scale variables take.  
In scaled 4-stop metroland the radial coordinate is the square root of the moment of inertia 
$\iota = \sqrt{I}$, the intuitive radius in mass-weighted configuration space, while for scaled 
triangleland this role is played by the moment of inertia $I$ itself.
Also, in scaled 4-stop metroland the $u^{\Delta}$ are the Cartesian components of the unit vector, 
while in scaled triangleland they are normalized Dragt-type coordinates \cite{Dragt}.  
The above quantities furthermore are good variables \cite{08II, AF, +tri, scaleQM} as regards 
kinematical quantization \cite{I84}, as well as identifiable as geometrically significant shape 
quantities.

In Sec 3, I consider RPM's in 

\noindent 
A) the indirect implementation of configurational relationalism in which the theories were originally 
conceived.  

\noindent 
B) In the {\it relationalspace approach}, i.e. following as the natural mechanics (in the sense of Jacobi 
and of Synge) constructed on the relationalspace geometry; this is 
a direct implementation of configurational relationalism which is available in 1- and 2-$d$.   
This approaches the identity of indescernibles directly: indiscernible configurations are never 
considered multiply.    

\noindent 
C) The {\it reduction scheme} involves eliminating A)'s auxiliary variables, and this is found 
to coincide with B), at least at the classical level.  
This approaches the identity of indiscernibles via their mathematical identification.  

\noindent D) I also consider rearranging Newtonian Mechanics by passing to absolute--relative split 
generalized coordinates, and compare the somewhat-related B), C).  

\noindent 
Finally, I also include some discussion of the physical interpretation of conserved quantities in RPM's.  
In Sec 4, I present 52 analogies between RPM's and GR (many of which are new to this paper).

In Sec 5, I account for differences in the form of Laplacian between the one I use for RPM's \cite{Banal} 
[which is quite in line with Quantum Cosmology \cite{DeWitt57, Magic, ConfOrder, Oporder}) and 
the one which is used in the Molecular Physics literature (which follows from scheme D)].  
This point is relevant both to the operator ordering problem in Quantum Gravity and Quantum Cosmology 
and as an `absolute imprint' contribution at the quantum level to the long-standing absolute versus 
relative motion debate. 
In Sec 6, I consider the further issue of whether RPM's configuration spaces have monopole issues, which 
is a significant preliminary consideration prior to being able to provide a detailed quantum study of 
scaled RPM's; this done, I can now present such quantum studies elsewhere \cite{scaleQM, 08III}.  
Corresponding gauge potentials (of Guichardet \cite{Guichardet} and various analogues thereto) are 
provided in the Appendix.

In Sec 7, I fix the aforementioned freedom in potential of RPM's via the Mechanics--Cosmology analogy, 
such that RPM's scale dynamics closely parallels that of known simple early universe cosmological models.    
This analogy is quite well-known in the case of late-universe dust models \cite{MMc,BGS03}, but I 
consider it in a wider sense that does not require similarness to the observed later universe but rather 
just mathematical parallels with simple plausible early universe cosmology models. 
This is a significant improvement in selection of potentials for RPM models, earlier ones having been 
chosen, on grounds of simplicity and/or good behaviour to be constant or multiple harmonic 
oscillator-like \cite{08I, 08II, AF, +tri, MGM}.  
The value of RPM toy models with such parallels is that their accompanying dynamics of small shape 
changes is, firstly, more tractable than GR Quantum Cosmology's accompanying dynamics of small 
inhomogeneities (e.g. Halliwell and Hawking's work \cite{HallHaw}), while also, secondly, unlike the 
consideration of small anisotropies in minisuperspace, being a bona fide local structure formation, 
which is relevant in e.g. semiclassical and records theoretic approaches to the Problem of Time.  
I support this in Sec 8 by considering simple classical solutions of the (approximately) separated-out 
scale part: which mechanics models cosmologically-standard models map to (and vice versa, to a certain 
extent).  
I apply this in Sec 9 to obtain approximate timestandards for the Semiclassical 
approach to the Problem of Time.  
Finally, the present paper's advances and clarifications make QM study possible for scaled 3- and 
4-stop metrolands \cite{scaleQM, SemiclIII} (\cite{MGM} has an outline for the former), and for scaled 
triangleland \cite{08III}; these papers will contain e.g. internal time, semiclassical approach, 
histories and records theory applications \cite{IT, MGM, SemiclIII, NOD, NOI}.

\section{Study of relational space}

\subsection{Relative space and Jacobi coordinates thereupon}  

A configuration space for N particles in dimension $d$ is $\fQ(\mN, $d$) = \mathbb{R}^{\sN d}$.  
I denote particle position coordinates by 

\noindent 
\{$\underline{q}^I$, $I$ = 1 to N\} and particle masses by 
$m_I$.   
Rendering absolute position irrelevant (e.g. by passing to any sort of relative coordinates) leaves one 
on a configuration space {\it relative space} = $\fR(\mn, \mbox{ } $d$) = \mathbb{R}^{\sn d}$ 
(for n = N -- 1).
The most obvious relative space coordinates formed from these are some basis set among the 
$\underline{r}^{IJ} = \underline{q}^J - \underline{q}^I$, however certain linear combinations of these 
--- {\it relative Jacobi coordinates} \cite{Marchal} --- prove to be more advantageous to use.  
I denote these by $\{\underline{R}^i$, $i = 1$ to n\}.
These are physically relative separations between {\sl clusters} of particles (see Fig 1 for the 
particular examples of these used in this paper), and their main advantage is that in them the kinetic 
metric is cast in diagonal form, and indeed looks just like for the $q^I$'s but involving one object 
less and with cluster masses $\mu_i$ in place of particle masses $m_I$ (c.f. Sec 3.2).
In fact, I use {\it mass-weighted} relative Jacobi coordinates $\underline{\iota}^i = \sqrt{\mu}_i 
\underline{R}^i$, their squares the partial moments of inertia $I^i = \mu_i\{R^i\}^2$ and `{\it 
normalized}' versions of both of these, $n^i$ and $N^i$ respectively (dividing by $\sqrt{I} \equiv 
\iota$ and $I$ respectively, where $I$ is the total moment of inertia of the system).

Moreover, for triangleland or 3-stop metroland, there are 3 permutations of the above, corresponding to 
following each of the particle clusterings \{1, 23\}, \{2, 31\} and \{3, 12\} respectively (Fig 1b). 
I use (a) as shorthand for \{a, bc\} where a, b, c form a cycle.  
On the other hand, quadrilateralland and 4-stop metroland both have both H-shaped and K-shaped Jacobi 
coordinates (Figs 1c and 1d).  
Each coordinate system `follows one clustering', so we need to use various coordinate systems (and this 
makes sense from a differential-geometric perspective).  
In 1-$d$, if we take abc to be distinct from cba (or abcd to be the same as dcba), our shapes are 
unoriented.  
In 2-$d$ the same applies if clockwise readings around the perimeter are taken to be distinct from 
anticlockwise ones.  
If these things are taken to be the same, one is considering oriented shapes.

{            \begin{figure}[ht]
\centering
\includegraphics[width=0.9\textwidth]{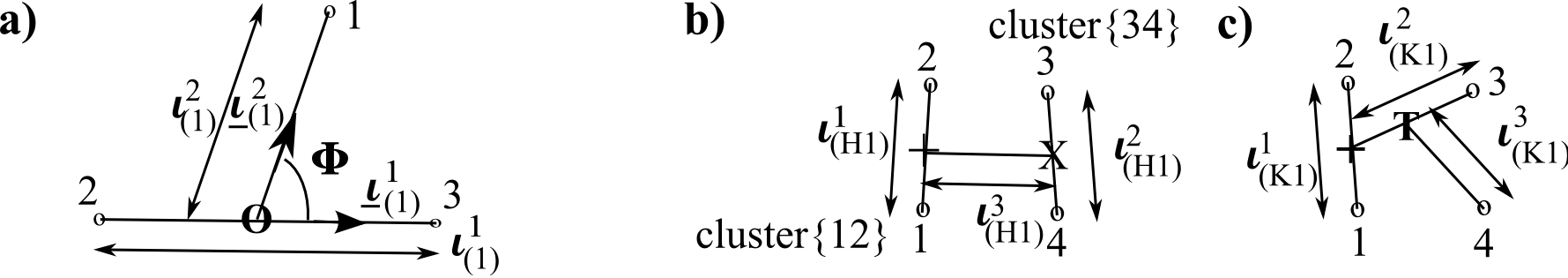}
\caption[Text der im Bilderverzeichnis auftaucht]{        \footnotesize{a) For 3 particles, one 
permutation of relative Jacobi coordinates is as indicated.  
As regards other permutations, I use $\iota^1_{(\sa)}$ and $\iota^2_{(\sa)}$ for the magnitudes of 
$\underline{\iota}^1_{(\sa)}$ and 
$\underline{\iota}^2_{(\sa)}$, while $\Phi_{(\sa)} = \mbox{arccos}\left(\underline{\iota}^1_{(\sa)}\cdot
{\underline{\iota}^2_{(\sa)}}/\iota^1_{(\sa)}\iota^2_{(\sa)}\right)$ is the angle between 
$\underline{\iota}^1_{(\sa)}$ and $\underline{\iota}^2_{(\sa)}$, and ${\cal R}_{(\sa)} = 
\iota^1_{(\sa)}/\iota^2_{(\sa)}$.

\noindent
O, X, + and * are the centres of mass of particles 2 and 3, 1 and 2, 3 and 4, and 
1, 2, 3 respectively.

\noindent b) One permutation of H-coordinates for 4 particles. 
As regards the other permutations, I use $\iota^i_{(\sH \sb)}$ where b takes 2, 3, 4 corresponding to 
clustering (1b, cd) `into pairs 1b and cd' where b, c, d form a cycle.

\noindent c) One permutation of K-coordinates for 4 particles.  
$\iota_{(\sK\sa)}$ where a takes 1, 2, 3, 4 corresponding to clusterings (a, bcd) where a, b, c and d 
form a cycle.
One can see that these coordinates form the shape of the letters H and K, hence these names. 
Finally, just flatten out each of the subfigures to obtain the 1-$d$ versions.] 
}        }
\label{Fig1}
\end{figure}            }

\subsection{More relational/more reduced configuration spaces}

If absolute scale is also to have no meaning, then one is left on a configuration space \cite{Kendall} 
{\it preshape space} = $\fP(\mn, d) = \fR(\mn, d)/$Dil (for Dil the dilational group); it is 
straightforward to see this to be $\mathbb{S}^{\sn d - 1}$.   
If absolute orientation (in the rotational sense) is also to have no meaning, then one is left on a 
configuration space {\it relational space} $\mbox{\Large r}(\mn, d) = \fR(\mn, d)/\mbox{Rot($d$)}$ 
(for Rot($d$) the $d$-dimensional rotation group).
If both of the above are to have no meaning, then one is left on \cite{Kendall} {\it shape space} = 
$\fS(\mn, d) = \fR(\mn, d)/\mbox{Rot}(d) \times \mbox{Dil}$.  
For $d$ = 1, being free of the rotations is trivial, so that $\fS(\mn, 1) = \fP(\mn, 1) = 
\mathbb{S}^{\sn - 1}$; this gives us that for 4 particles on a line the configuration space is 
$\mathbb{S}^2$.
For $d$ = 2, $\fS(\mn, 2) = \fR(\mn, 2)/SO(2) \times \mbox{Dil} = \mathbb{S}^{2\sn - 1}/U(1) = 
\mathbb{CP}^{\sn - 1}$ \cite{Kendall, FORD}, while the well-known result $\mathbb{CP}^{1} = 
\mathbb{S}^2$ then gives that $\fS(3, 2) = \mathbb{S}^2$.  
[The situation rapidly gets more complicated for increasing numbers of particles in 3-$d$.] 
The above also assumes each shape and its mirror image are considered distinct (the `plain' 
choice of shapes); if instead one were to identify each shape and its mirror image, one 
would be considering the `oriented' shapes, and one would get quotients of all of the above spaces by 
$\mathbb{Z}_2$. 
Among these, the real projective spaces $\mathbb{RP}^k = \mathbb{S}^k/\{\mathbb{Z}_2$ with inversive 
action\} are well-known.
I use an extra $\fS$ to denote the configuration spaces in this case, e.g. $\fS\fS$(n, $d$).

Next, consider these spaces at the level of Riemannian geometry (which gives a `distance' between 
relative configurations/shapes from the Riemannian line element -- a useful structure for timeless 
approaches such as the na\"{\i}ve Schr\"{o}dinger interpretation or records theory).  
Denote shape space line element by $\d s^2$; the corresponding Riemannian space is $\langle \sigma , 
\bm \rangle$ for $\sigma$ a topological surface and $\bm$ the Riemannian metric associated with 
$\d s^2$.  
For $\mathbb{S}^{\sn - 1}$, it is the line element
\beq
\d s^2_{\sN-\sss\st\so\sp\,\sS\sR\sP\sM} = \sum \mbox{}_{\mbox{}_{\mbox{\scriptsize $\barr = 1$}}}^{\sn - 1}
\prod\mbox{}_{\mbox{}_{\mbox{\scriptsize $\hat{p} = 1$}}}^{\barr - 1}\,\mbox{sin}^2\Theta_{\hat{p}}
\d{\Theta}_{\barr}^2 \mbox{ } 
\eeq
in standard (hyper)spherical coordinates \{$\Theta_{\bar{r}}$, $\bar{r} = 1$ to $\n - 1$\}
For $\mathbb{CP}^{\sn - 1}$, it is the Fubini--Study line element 
\beq
\d s^2 = 
\{\{1 + |{\cal Z}|_{\sc}^2\}|\d {\cal Z}|_{\sc}^2 - ({\cal Z} , \d \overline{{\cal Z}})^2\}/
 {\{1 + |{\cal Z}|_{\sc}^2\}^2}  \mbox{ } .
\label{FS} 
\eeq
Here, $|{\cal Z}|_{\sc}^2 = \sum_{\bar{r}}|{\cal Z}^{\bar{r}}|^2$, $( \mbox{ } , \mbox{ } )_{\sc}$ is 
the corresponding inner product, \{${\cal Z}^{\bar{r}}$, $\bar{r} = 1$ to $\mn$ -- 1\} are the 
inhomogeneous coordinates (independent set of ratios of $z^{i} = {\cal R}^{i}\mbox{exp}({i\Theta^{i}})$ 
for ${\cal R}^{i}$, $\Theta^{i}$ the polar presentation of Jacobi coordinates), the overline denotes 
complex conjugate and $|\mbox{ }|$ the complex modulus.
The unoriented 3-stop metroland circle case has coordinate range 0 to $2\pi$, while the oriented one has 
0 to $\pi$.  
One can use a distinct double angle coordinate to make this 0 to $2\pi$ again.

The general sphere notation and specific notations for subcases I use is as follows. 
$\alpha$ is an azimuthal angle (range 0 to $\pi$) and $\chi$ is a polar angle (range 0 to 2$\pi$).  
Then the spherical line element is 
\beq
\d s^2 = \d\alpha^2 + \mbox{sin}^2\alpha\,\d\chi^2 \mbox{ } .  
\eeq
[In the case of the sphere in actual space, which sometimes makes for a useful analogy, I denote 
$\alpha$ by $\theta_{\sss\sp}$ and $\chi$ by $\phi_{\sss\sp}$)].  
For the 4-stop metroland line element, use $\alpha \longrightarrow \theta$, $\chi \longrightarrow \phi$.  
The oriented case has the half-range 0 to $\pi/2$ for $\theta$, with inversive identification on the 
equator that makes it $\mathbb{RP}^2$.  
For the triangleland line element, use $\alpha \longrightarrow \Theta = 2\,\mbox{arctan}\,{\cal R}$ for 
${\cal R} = |\underline{\iota}_1/\underline{\iota}_2|$ and $\chi \longrightarrow \Phi$.  
The oriented case has the half-range 0 to $\pi/2$ for $\theta$, with reflective identification about the 
equator that makes it a hemisphere with boundary rather than a $\mathbb{RP}^2$.  
Thus the oriented 4-stop metroland and triangleland are mathematically distinct.  
The point of this is that, while we have the same mathematics as for the standard situation of a sphere 
in space, we now have this tractable mathematics alongside a new interpretation that makes this 
appropriate as a whole-universe toy model (which is then highly soluble in many different ways).

\subsection{Relational space as the cone over shape space}

Relational space $\bigr(\mN, d)$ can be viewed as the cone\footnote{\noindent Such  
a notion of cone is thought to have first appeared in the works of Lema\^{\i}tre and Deprit--Delie 
\cite{LemDD}.
(It is also used in 3-body work in classical mechanics such as \cite{MoeckelQSaariMont, Hsiang, Mont98}.)}
over shape space, $C(\fS(\mN, d))$.
At the topological level, for C(X) to be a cone over some topological manifold X, 
\beq
\mbox{C(X) = X $\times$ [0, $\infty$)/\mbox{ }$\widetilde{\mbox{ }}$} \mbox{ } , 
\eeq
where $\widetilde{\mbox{ }}$ means that all points of the form (p $\in$ X, 0 $\in [0, \infty)$ ) are 
`squashed' or identified to a single point termed the {\it cone point}, and denoted by 0. 
At the level of Riemannian geometry (see e.g. \cite{Mont98, Hsiang}), a cone C(X) over a 
Riemannian space X possesses a) the above topological structure and b) a Riemannian line 
element given by 
\beq
\d S^2 = \d \rho^2 + \rho^2\d s^2 \mbox{ } ,
\eeq
where $\d s^2$ the line element of X itself and $\rho$ is a suitable `radial variable' 
that parametrizes the [0, $\infty$), which is the distance from the cone point.  
This metric is smooth everywhere except (possibly) at the troublesome cone point.

\noindent Note I) The everyday-life cone can indeed be viewed as a simple example of this construction, 
using X = (a piece of) $\mathbb{S}^1$.  

\noindent Note II) At the Riemannian level, we have a notion of distance and hence (for sufficiently 
nontrivial dimension) of angle, so that one can talk in terms of deficit angle.  
C($\mathbb{S}^1$) itself has no deficit angle, while using a $p$-radian piece entails a deficit angle of 
$2\pi - p$.  
The presence of deficit angle, in turn, gives issues about `conical singularities', e.g. \cite{KS}.   

\noindent Notes I) and II) straightforwardly generalize to $\mathbb{S}^k$ (via `deficit solid 
angle').

I would like to push the above definition as far as possible toward cases in which X is a stratified 
manifold.  
Cones are examples of {\it orbifolds}, though these have more structure defined, as follows.  
Paralleling how real and complex manifolds are defined in terms of charts to $\mathbb{R}^k$ and 
$\mathbb{C}^k$, real and complex orbifolds are defined in terms of charts to quotients of these by a 
group.  
Another means of formulating the kind of mathematics arising in the relational program is 
\beq 
\mbox{C(X) for X = $R/G$} \mbox{ } , 
\eeq 
i.e. the cone over a topological space $R$ quotiented by a group $G$ (which may be continuous, discrete 
or a mixture of both).
E.g. C($\fS(\mn, d)$) = C($\mathbb{R}^{\sn d}/SO(d) \times$ Dil), including $\mC(\fS(\mn, 1)) = 
\mC(\mathbb{S}^{\sn - 1})$ and $\mC(\fS(\mn, 2)) = \mC(\mathbb{CP}^{\sn - 1})$, C($\fS\fS(\mn,\d)$) = 
C($\mathbb{R}^{\sn d}/SO(d) \times$ Dil $\times \mathbb{Z}_2$), including C($\fS\fS(\mn, 1)$) = 
C($\mathbb{S}^{\sn - 1}/\mathbb{Z}_2$) and C($\fS\fS(\mN, 2)$) = 
C($\mathbb{CP}^{\sn - 1}/\mathbb{Z}_2$), and further examples involving bigger discrete groups in 
models with particle indistinguishibility.    
A partial answer to whether cones are still definable over such is that Acharya, Atiyah and Witten 
\cite{W1W3} do talk about cones over weighted projective spaces, which can, at least in some 
cases, possess orbifold singularities.

Finally, by the shape--scale split, shape space is both the entirety of the reduced configuration space 
for the scalefree theory and the shape part of the shape-scale split of the corresponding scaled theory.

\subsection{This paper's examples' cones are straightforward}

C($\fS$(n, 1)) = C($\mathbb{S}^{\sn - 1}$) = $\mathbb{R}^{\sn}$ using just elementary results (c.f. e.g. 
\cite{Kendall, Rotman}).   
C($\fS$(3, 2)) = C($\mathbb{CP}^{1}$) = C($\mathbb{S}^{2}$ of radius 1/2) = $\mathbb{R}^{\sn}$ (see e.g. 
\cite{Hsiang}) up to a conformal factor at the metric level (Sec 2.6) which can be `passed' to the 
potential, as in Sec 3.  
C($\fS\fS$(n, 1)) = C($\mathbb{S}^{\sn - 1}/\mathbb{Z}_2$) = $\mathbb{R}^{\sn}_+$, the half-space i.e. 
half of the possible generalized deficit angle.  
C($\fS\fS$(3, 2)) = C($\mathbb{S}^2/\mathbb{Z}_2$) = $\mathbb{R}^3_+$.   
N.B. this is {\sl not} $\mathbb{RP}^2$ but rather the sphere with reflective rather than inversive 
$\mathbb{Z}_2$ symmetry about the equator; one can think of this space loosely as a `half-onion'.
These oriented examples have additional `edge issues'. 
In 3-stop metroland, the identification coincides with the gluing in constructing the `everyday cone'.  
Oriented 4-stop metroland is $\mathbb{RP}^2$. 
Moreover, one often has to exclude the collinear plane for this and for oriented triangleland.

\subsection{Topological structure of cones over shape spaces.} 

Cones have a tendency to be straightforward from the topological point of view.  
Certainly the main particular examples of this paper, which reduce to $\mathbb{R}^{\sn}$ (see e.g. 
\cite{Hsiang} as regards unoriented triangleand being homeomorphic to $\mathbb{R}^{3}$), 
and $\mathbb{R}^{\sn}_+$, are topologically straightforward.  
Also, C(X) is contractible (pp. 23-24 \cite{Rotman}), and so has the same homotopy type as the point.
Furthermore, cones are acyclic and as such have no nontrivial homology groups (pp. 43-46 of 
\cite{Munkres}). 
However, some applications require excision of the cone point, giving punctured cones.    
%
%
As regards homotopies, this effectively amounts to a return to the shape space by means of a retract.   
The situation with cohomologies is also straightforwardly related to that of the shape space X 
\cite{KW06}.
Then results for $\mathbb{S}^k$, $\mathbb{CP}^k$ shape space topology are tabulated in e.g. \cite{FORD}, 
$\mathbb{RP}^k$ counterparts of these can be patched together from e.g. 
\cite{HatcherSpanierGreenbergandHarper}, and the hemisphere is also topologically standard.

Singular potentials can require exision of further points, namely the collision set $\Sigma$ (i.e. 
non-maximal as well as maximal collisions); these are exemplified in Sec 3 for this paper's models. 
The potentials in my model to date have been benign -- of the $r^2$ type, for which this does not occur.    
However Sec 7--8's cosmological motivation does bring in potentials that are less benign in this sense.

To demonstrate that this is indeed capable of altering -- and substantially complicating -- the 
topology, I give the following example. 
The topologically trivial $\fQ(\mN, d) = \mathbb{R}^{\sn d}$, upon excision, picks up 
$\pi_1(\fQ(\mN, d)/\Sigma)$ = \{coloured braid group\}, the colouring referring to particle 
distinguishability.  
(That these are isomorphic is clear given that the particles are assumed distinguishable and the orbits 
can wind around each of the binary collisions in whatever order but not intersect with them, i.e. the 
definition of a coloured braid with each colour corresponding to a distinguishable particle.  
In fact, this structure was first found in this context by Hurwitz in 1891 \cite{Hurwitz}, thus 
preceding the realization that such as structure is a braid group by Artin in 1925 \cite{Artin}; these 
results were first put together in \cite{FN62}.
See e.g. \cite{Birman} for a review and updates on the theory of the braid group.)
\cite{Mont98} takes this further for 3 particles in 2-$d$; see also \cite{T90AG} for further 
applications.

\subsection{Metric structure of cones over shape spaces}  

The above shape analysis can be uplifted to above to ERPM, alongside a further size variable, by the 
cone construction/ shape--scale split.   
This further variable can be taken to be $\iota = \sqrt{I}$ for N-stop metroland and $I$ for 
triangleland.
$\iota = \sqrt{I}$ is termed the {\it hyperradius}, and its use apparently dates back to Jacobi 
\cite{ACG86}; it was used in QM at least as far back as the 1950's by Fock \cite{Fock} and by Morse and Feschbach 
\cite{MFII}; see also e.g. \cite{Smith62, Mont96, Hsiang}.
$I$ is used e.g. in \cite{Dragt, Iwai, LR} though these are for $\theta$ running over half of the range 
most often used in the present paper (but there is an analogous use of $I$ in the case 
involving the whole range too).

The unoriented N-stop metroland case is straightforward as C($\mathbb{S}^{\sn - 1}$) is  
$\mathbb{R}^{\sn}$  (c.f. e.g. \cite{PP87}).
The oriented version can be viewed as having a deficit angle of $\pi$ for 3 particles and a deficit 
solid angle of 2$\pi$ for 4 particles.  
\beq
\d S^2_{3-\sss\st\so\sp\,\sE\sR\sP\sM} = \d\iota^2 + \iota^2\d\varphi^2 \mbox{ } , 
\eeq
\beq
\d S^2_{4-\sss\st\so\sp\,\sE\sR\sP\sM} = \d\iota^2 + \iota^2\{\d\theta^2 + \mbox{sin}^2\theta\d\phi^2\} 
\mbox{ } , 
\eeq
while, for scaled triangleland, as the shape space is the sphere of radius 1/2,
\beq
\d S^2_{\triangle\,\sE\sR\sP\sM} = \d\iota^2 + \{\iota^2/4\}\{\d\Theta^2 + \mbox{sin}^2\Theta\d\Phi^2\}
\mbox{ } , 
\eeq
which inconvenience in coordinate ranges can be overcome\footnote{See 
\cite{J83} for a distinct way of doing this.}  
by using, instead, $I$ as the radial variable, 
\beq
\d S^2_{\triangle\,\sE\sR\sP\sM} = \{1/4I\}\{\d I^2 + I^2\{\d\Theta^2 + \mbox{sin}^2\Theta\d\Phi^2\}\} \mbox{ } ,
\eeq 
which metric itself is conformally flat; the flat metric itself is 
\beq
\d S^2_{\sf\sll\sa\st} = \d I^2 + I^2\{\d\Theta^2 + \mbox{sin}^2\Theta\d\Phi^2\}
\eeq
(spherical polar coordinates with $I$ as radial variable). 
This using of $I$ as radial variable: the start of significant differences between the triangleland 
and 4-stop metroland spheres and half-spheres.
In the oriented case, one can also use a double angle variable running over the usual range of angles
e.g. \cite{Smith62, LMAC98}.

Some authors such as Iwai \cite{Iwai} consider the 3-$d$ case, for which it is the oriented version 
that occurs naturally.  
Also note that indistinguishability issues also cut down on the exent of the configuration space.  
There are also conical singularity issues with such spaces.  
These could lead to having to exise the cone point.

I leave details of the metric and topological structure of C($\mathbb{CP}^{\sn}$) with $\mn > 1$, 
C($\mathbb{RP}^{\sn}$) with $\mn > 2$ and C($\mathbb{CP}^{\sn}/\mathbb{Z}_2$) with $\mn > 1$ as 
problems for a future occasion.  
Minus the cone point, unoriented triangleland is diffeomorphic to $\mathbb{R}^3$ \cite{Hsiang}.  
Is the version I consider in \cite{08I}.  
If one keeps the $1/4I$ factor, the geometry is curved, with a curvature singularity at 0, and, 
obviously, conformally flat where the conformal transformation is defined (i.e. elsewere than 0).  
In any case the $E/I$ `potential term' in  \cite{08I, 08III} is singular there too.

\subsection{Tessellation by the physical interpretation}

This is useful as regards reading off the physical interpretation: classical trajectories can be 
interpreted as paths thereupon, and classical potentials and QM probability density 
functions as height functions thereover.

{            \begin{figure}[ht]
\centering
\includegraphics[width=0.5\textwidth]{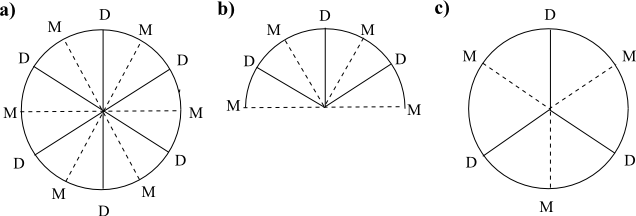}
\caption[Text der im Bilderverzeichnis auftaucht]{        \footnotesize{Tessellation of 3-stop 
metroland's configuration spaces.}        }
\label{Fig2}
\end{figure}            }

I give here the scaled 3-stop metroland case as an example; the scaled 4-stop metroland and triangleland 
cases following straightforwardly from more intricate examples of tessellations presented elsewhere 
\cite{AF, +tri}.  
For 3-stop metroland \cite{06I}, in the oriented case (the rim of Fig 2a), the shape space is the 
circle with 6 regularly-spaced double collision (D) points on it, and, somewhat less significantly 6 
more points ({\it mergers}, M: where the centre of mass of two of the particles coincides with the third 
particle) halfway between each two adjacent D's thus forming an hour-marked clock face.  
This has symmetry group\footnote{I use 
$\mathbb{D}_p$ for the dihaedral group of order 2$p$, and $S_p$ for the permutation group of $p$ objects, 
of order $p!$}
$\mathbb{D}_6 \mbox{ } \widetilde{=} \mbox{ } \mathbb{D}_3 \times \mathbb{Z}_2 \mbox{ } \widetilde{=} 
\mbox{ } S_3 \times \mathbb{Z}_2$ i.e. the freedom to permute particle labels and to ascribe an 
orientation, however as M's and D's are clearly physically distinct, the tessellation labelled by these 
has its symmetry reduced to $\mathbb{D}_3$.  
The 6 D's pick out 3 preferred axes, and the 6 M's pick out 3 further preferred axes.     
The oriented counterpart (the half-rim in Fig 2b) has 3 D's and 3 M's and the symmetry group 
$\mathbb{D}_3 \mbox{ } \widetilde{=} \mbox{ } S_3$; this can also be represented using double angles 
as the rim of a whole pie (Fig 2c).    
Now each D is opposite an M, so the DM pairs pick out 3 preferred axes. 
Each of these corresponds to one of the 3 permutations of Jacobi coordinates; this is also the case for 
the 3 axes picked out by the D's in the unoriented case.   
The polar angle $\varphi^{(\sa)}$ about each of these axes is then natural for the study of the 
clustering corresponding to that choice of Jacobi coordinates, \{a, bc\}.  
The relational spaces for the scaled 3-stop metroland theories are then the cones over these decorated 
shape spaces,  i.e. (the infinite extension of) Fig 2a)'s pie of 12 slices with 6 D half-lines and 6 M 
half-lines in the unoriented case and likewise Fig 2b)'s half-pie of 6 slices (or Fig 2c)'s pie of 3 
slices) with 3 D half-lines and 3 M half-lines in the oriented case.   
All of these emanate from the triple collision at the cone point, 0.

A number of useful propositions concerning regions of the shape space then involve arcs, e.g. small 
arcs around decorated points such as $|\varphi^{(\sa)}| \leq \epsilon$ about a particular D or $\pi/6 - 
\epsilon \leq \varphi^{(\sa)} \leq \pi/6 + \epsilon$ about a particular M in the unoriented case.  
For the relationalspace, propositions can now concern e.g. a disc $\iota \leq \varepsilon$ 
around the maximal collision, or sectors (parametrized by a $\varphi^{(\sa)}$) around or far from e.g. 
certain of the decorated lines.  
Regions corresponding to  
more complicated propositions can then be built up from discs and sectors under the operations of 
union, intersection and negation (e.g. arcs of annulus).  
Note 1) that inverse power law potentials are singular on D's so \cite{PP87} classical motion and QM 
wavefunctions effectively split up into $\pi/3$ wedges.
The above sort of approximate notions of shape are in the spirit of those used in e.g. Kendall et al. 
\cite{Kendall}, and we make use of the corresponding configuration space regions in na\"{\i}ve 
Schr\"{o}dinger approach calculations in \cite{AF, +tri, scaleQM, 08III} and as regards in which 
regions semiclassicality holds.  
The corresponding tessellations for scalefree 4-stop metroland and triangleland are in 
\cite{AF, +tri}; the scaled counterparts of these are just the cones over these decorated spheres, 
and useful propositions now correspond to a number of standard solid regions such as solid sectors, 
solid (ordinary) cones and (pieces of) solid shells.  
For negative power-law potentials, edges representing collisions can become singular, so that the 
subsets of the configuration space can/must be treated separately.

\subsection{Cartesian map versus Dragt map}

For 3-stop metroland, use polar angles for SRPM/the shape space within ERPM, and these alongside a 
radial coordinate to make plane polar coordinates for ERPM's relational space.  
In the oriented counterpart, one can consider a half-circular arc/half space or use a double angle 
coordinate to get a whole circle/whole plane.  
Shape quantities for this are $n^1_{(\sa)} = \mbox{cos}\,\varphi^{(\sa)}$, $n^2_{(\sa)} = \mbox{sin}\,
\varphi^{(\sa)}$, so $\{n^1_{(\sa)}\}^2 + \{n^2_{(\sa)}\}^2 = 1$; also, inverting, $\varphi^{(\sa)} = 
\mbox{arctan}(n^1_{(\sa)}/n^2_{(\sa)})$.  
Both the above and the below should be linked to kinematic quantization, pointing out that inclusion of 
a radial variable is also a straightforward and well-known procedure, so passage from SRPM to ERPM in 
shape-scale split variables is no problem in this regard.  
%

Working at the level of a general joint treatment, one can recast the above overarching mathematical 
solution in terms of three variables $u^{\Delta}$ such that $\sum_{\Delta = 1}^3\{u^{\Delta}\}^2 = 1$.
These describe a Euclidean 3-space that surrounds the sphere.  
It is then often convenient to use $u_x$, $u_y$ and $u_z$ for the components of $u^{\Delta}$.    
The $u^{\Delta}$ are related to the $\alpha$ and $\chi$ through being the components of the 
corresponding unit Cartesian vector in spherical polar coordinates:
\beq
u_x = \mbox{sin}\,\alpha\,\mbox{cos}\,\chi\, \mbox{ } , \mbox{ } \mbox{ }
u_y = \mbox{sin}\,\alpha\,\mbox{sin}\,\chi\, \mbox{ } , \mbox{ } \mbox{ }  
u_z = \mbox{cos}\,\alpha \mbox{ } . 
\eeq 
For $\mathbb{S}^2$ in `actual space', the $\mathbb{R}^3$ {\sl is} `actual space' with the physical radius $r$ in 
the role of the radial coordinate, while for 4-stop metroland and triangleland RPM's, it is relational 
space with as radius, respectively, the `natural' $\iota$ of mass-weighted space, and $I = \iota^2$ 
(because triangleland's shape sphere arises from $\mathbb{CP}^1$, which gives its natural radius an 
unusual factor of 1/2, which is absorbed by the coordinate transformation to radial variable $I$).  
This makes triangleland quite unlike 4-stop metroland or actual space as regards the physical meaning 
of its $u^{\Delta}$.  
In the latter cases, they are simply the Cartesian components $x^{\alpha}$ and $n^i$ (trivial 
`Cartesian maps').  
However, for triangleland, the $u^{\Delta}$ are related to the configuration space's coordinates, 
rather, by the `Dragt map' \cite{Dragt}:\footnote{See 
\cite{Gronwall,Smith62} for earlier literature and e.g. \cite{MTAqui93ML99,Iwai,PP87,LR} for some 
applications.
There also being some slight differences here between 2-$d$ and 3-$d$, oriented and unoriented 
triangles, and the scaled and scalefree cases, I refer to the coordinates I use as of `Dragt-type'.}
%
\beq
\mbox{dra}^{(\sa)}_x = \mbox{sin}\,\Theta_{(\sa)}\,\mbox{cos}\,\Phi_{(\sa)} = 
2n^1_{(\sa)}n^2_{(\sa)}\,\mbox{cos}\,\Phi_{(\sa)} 
\mbox{ } ,
\label{dragt1}
\eeq
\beq
\mbox{dra}^{(\sa)}_y = \mbox{sin}\,\Theta_{(\sa)}\,\mbox{sin}\,\Phi_{(\sa)} = 
2n^1_{(\sa)}n^2_{(\sa)}\,\mbox{sin}\,\Phi_{(\sa)} 
\mbox{ } ,
\label{dragt2}
\eeq
\beq
\mbox{dra}^{(\sa)}_z = \mbox{cos}\,\Theta_{(\sa)} = 
N^2_{(\sa)} - N^1_{(\sa)} 
\mbox{ } .   
\label{dragt3}
\eeq
In these depending on squared quantities, one can see consequences of $\{\iota\}^2 = I$ and not $\iota$ 
being the radial coordinate.
So one goes from ($\alpha, \chi$) to $u^{\Delta}$ $\Delta$ = 1 to 3, and then, for 4-stop metroland, to 
$n^i_{(\sH \sb)}$ via the Cartesian map, and, for triangleland to $\mbox{dra}^{\Delta}_{(\sa)}$ via the 
Dragt map.  
This can be used to check 4-stop metroland and triangleland results against each other, and to extend 
what has been done in one of these models to the other (at least when both problems remain analogous, as 
the general study here of HO-like potentials does eventually break the analogy).

N.B. the usual Dragt coordinates are related to ours by $\mbox{Dra}^{\Delta} = I\,\mbox{dra}^{\Delta}$; 
moreover in the literature it is usually the oriented half-space case for which these are presented.

Also, moving between different clusterings involves linear transformations $u^{\Delta} = 
{D^{\Delta}}_{\Gamma}u^{\Gamma}$, termed `democracy transformations' in e.g. \cite{LR95}.  
The present paper's notion of `clustering invariant' thus coincides with the Theoretical Molecular 
Physics literature's notion of `democracy invariant'.

\subsection{Shape quantities}

The $u^{\Delta}$ are useful variables (e.g. as regards kinematical quantization \cite{I84}).  
I next discuss their meaning for 4-stop metroland and triangleland.
{\footnotesize 
\begin{tabbing}
\underline{Geometrical Quantity} \hspace{0.05in}          \= 
\underline{3-stop metroland meaning} \hspace{0.05in}      \=
\underline{4-stop metroland H-meaning} \hspace{0.05in}    \=
\underline{4-stop metroland K-meaning} \hspace{0.05in}    \=
\underline{triangleland meaning} \hspace{0.3in}          \\

$u^1$                               \> 
$n^1_{(\sa)}$ = RelSize(bc)           \>  
$n^1_{(\sH \sb)}$ = RelSize(1b)     \>  
$n^1_{(\sK \sa)}$ = RelSize(ab)     \>
dra$^1_{(\sa)}$ = Aniso(a)        \\

$u^2$                               \> 
$n^2_{(\sa)}$ = RelSize(bc, a)        \>  
$n^2_{(\sH \sb)}$ = RelSize(cd)     \>
$n^2_{(\sK \sa)}$ = RelSize(ab,c)   \>  
dra$^2_{(\sa)}$ = TetraArea       \\

$u^3$                               \> 
\hspace{0.3in}  --                  \>   
$n^3_{(\sH \sb)}$  = RelSize(1b,cd) \>  
$n^3_{(\sK \sa)}$  = RelSize(abc,d) \>  
dra$^3_{(\sa)}$ = Ellip(a)        \\

-----------------------------------\>
------------------------------------------\>
--------------------------------------------\>
--------------------------------------------\>
-----------------------------\\

Linear functions   \hspace{8in}     \>\>\>                    
  
$N^1$ = Tall(a) \\

thereof   
\hspace{2.05in} \> 
\hspace{0.4in}  -- \> 
\hspace{0.4in}  -- \> 
\hspace{0.4in}  -- \>  
$N^2$ = Flat(a) \\ 

\end{tabbing}                                 }

\noindent 
Scaled 3- and 4-stop metroland's Cartesian components are all cluster-dependent ratio quantities 
concerning how large a given cluster is relative to the whole model or how well-separated the two 
clusters are in the latter case that has enough particles to build up such a quantity.

For 3-stop metroland \cite{AF}, $n^1_{(\sH \sb)}$ is a measure of how large the universe is relative to 
cluster bc, and $n^2_{(\sH \sb)}$ is a measure of how large the universe is relative to the separation 
between cluster bc and particle a.  
Thus I term these respectively RelSize(bc) and RelSize(bc, a).
For 4-stop metroland \cite{AF}, $n^3_{(\sH \sb)} =$ RelSize(1b,cd) is a measure of how large the 
universe is relative to the separation between clusters 1b and cd, $n^1_{(\sH \sb)} =$ RelSize(1b)
is a measure of how large the universe is relative to cluster 1b, and $n^2_{(\sH \sb)} = $ RelSize(cd) 
is a measure of how large the universe is relative to cluster cd.  
One can also construct K (rather than H) analogues of the above notions for use in following triple clusters: 
$n^3_{(\sK \sa)}$ = RelSize(abc, d), a measure of how large the universe is relative to the separation 
between cluster abc and particle d, $n^2_{(\sK \sa)}$ = RelSize(ab,c), a measure of how large the 
universe is relative to the separation between subcluster ab and particle c, and $n_2^{(\sH \sb)}$ = 
RelSize(ab), a measure of how large the universe is relative to cluster ab.  
Note that these have a more symmetric meaning in H coordinates than in K coordinates or for 3 particles.

For triangleland, assuming equal masses for simplicity (see \cite{+tri} for elsewise),
$\mbox{dra}^{(\sa)}_z$ is the `ellipticity' Ellip(a) = $N^2_{(\sa)} - N^1_{(\sa)}$ of the two 
`normalized' partial moments of inertia involved in the (a)-clustering. 
This is a pure ratio (rather than relative angle) quantity.  
It is closely related to the tallness quantity Tall(a) = $N^2_{(\sa)}$ -- how tall the triangle is with 
respect to the (a)-clustering, and Flat(a) = $N^1_{(\sa)}$, which is likewise a flatness quantity: 
Tall(a) = $\{1 + \mbox{Ellip(a)}\}/2$, $\mbox{Flat(a)} = \{1 -  \mbox{Ellip(a)}\}/2$.
$\Theta_{(\sa)}$ itself can also be considered as a ratio variable.  
One can view dra$^{(\sa)}_x$ as a measure of `anisoscelesness' Aniso(a) (i.e. departure from (a)'s notion of 
isoscelesness; c.f. `anisotropy' as a departure from isotropy), in the sense explained in \cite{+tri}.  
One can likewise view dra$^{(\sa)}_y$ as a measure of noncollinearity; moreover it is actually 
clustering-independent alias a `democracy invariant' \cite{Zick, ACG86, LR95} and furthermore equal to 
four times the area of the triangle per unit $I$ in mass-weighted space, and so I term it TetraArea.  
In contrast, 4-stop metroland has no democratic invariant.  
For extension to bigger particle numbers, see \cite{QShape}.

In the ERPM case, by the shape--scale split, shape quantities remain significant, now alongside size 
quantities $\iota \equiv$ Size and $I \equiv$  SIZE.  
Finally, note that I suppress clustering labels from now on (for 4-stop metroland, what I consider 
is one of the H-clusterings).

\section{Various formulations of RPM's}

\subsection{Temporally Relational Actions}

ERPM follows from temporal relationalism implementing reparametrization-invariant product-type actions,  
\beq
\fS^{\sE\sR\sP\sM} = 2\int\d\lambda\sqrt{T\{E - V\}} \mbox{ } .  
\label{prod}
\eeq
Here, $T$ is the kinetic term (I also use $K = 2T$), $V$ is the potential term, and $E$ is the total energy.
Such actions can either incorporate configurational relationalism indirectly via auxiliary 
$G$-variables, or directly by being constructed to be explicitly $G$-invariant (see below for details 
of both).   
Reducing the former gives the latter, at least at the classical level.  
Product-type action admits an obvious `banal conformal invariance' \cite{Banal}, permitting access from 
conformally flat metric to flat metric at cost of placing weighting on $V$ and $E$.  
Such product-type actions possess the `banal invariance' under $T \longrightarrow \Omega^2T$, $E - V 
\longrightarrow \Omega^{-2}\{E - V\}$.   
Then the derivative with respect to the emergent time, $* \equiv \d/\d t = \sqrt{\{E - V\}/T}
\dot{\mbox{ }}$ scales as $* \longrightarrow \Omega^{-2}*$.  
I use various banal representations in this paper: the trivial one ($\Omega = 1$) the $t$ for which I will 
refer to as `emergent Jacobi--Barbour--Bertotti time', the flat one for triangleland, which I denote 
by checking, for which $\Omega = 4I$, so $\check{T} = 4I T$, $\check{E} - \check{V} = \{E - V\}/4I$ and  
\beq
\check{*} \equiv \sqrt{\{\check{E} - \check{V}\}/{\check{T}}} \mbox{ } \dot{\mbox{}} = 
\{4\mI\}^{-1}\sqrt{{E - V}/{T}} \mbox{ } \dot{\mbox{}} = \{4\mI\}^{-1} * \mbox{ } ,  
\label{checkstardef}
\eeq
and ones for which shape space comes out geometrically naturally (with $\Omega = 1/\iota$ for 1-$d$ 
and the $\mathbb{CP}^{\sn - 1}$ presentation of 2-$d$, and with $\Omega = 1/I$ for the $\mathbb{S}^2$ 
presentation of triangleland).

\subsection{Indirect presentations of ERPM}

ERPM in its traditional indirect presentation (c.f. \cite{BB82} though the below additionally applies 
the cyclic velocities formulation \cite{B03, ABFO, FEPI} in terms of particle postion coordinates is 
(\ref{prod}) with\footnote{I 
use $||\mbox{ }||_{\tbg}$ for the norm involving the metric $\bg$ and $(\mbox{ },\mbox{ })$ for the 
corresponding inner product; in the case in which the metric is the ordinary flat space one, I use the 
unadorned $||\mbox{ }||$ or $(\mbox{ },\mbox{ })$.}
\beq
T = \sum\mbox{}_{_{\mbox{\scriptsize I = 1}}}^{\sN} m_I||\Circ_{\sA,\sB}\underline{q}^I||^2/2 \mbox{ } , \mbox{ } \mbox{ } 
\Circ_{\sA,\sB}q^{\alpha I} \equiv 
\dot{\underline{q}}^{I} - \dot{\underline{\mA}} - \dot{\underline{\mB}} \cr \underline{q}^{I} \mbox{ } .  
\eeq
The latter are arbitrary $G$-frame corrected velocities [here with $G$ the Euclidean group Eucl(N, $d$) of 
translations and rotations; \underline{A} and \underline{B} are translational and rotational 
auxiliaries].
The conjugate momenta are then
\beq
\underline{p}_I = m_I\delta_{IJ} \Star_{\sA,\sB}\underline{q}^{J} 
\mbox{ } , \mbox{ } \mbox{ }
\Star_{\sA,\sB}q^{\alpha I} \equiv 
\Star{\underline{q}}^{I} - \Star{\underline{\mA}} - \Star{\underline{\mB}} \cr \underline{q}^{I} 
\mbox{ } . 
\eeq
These momenta obey as a primary constraint that is quadratic and not linear in the momenta the `energy 
constraint' 
\beq
\ttQ \equiv \sum\mbox{}_{_{\mbox{\scriptsize I = 1}}}^{\sN} \, p_I\mbox{}^2/2m_I  + V = E \mbox{ } , 
\label{EnCo}
\eeq
the left hand side of which is also the restricted \cite{Dirac} Hamiltonian for the system. 
Variation with respect to $\underline{A}$ and $\underline{B}$ gives as secondary constraints linear in 
the momenta constraints 
\beq
\underline{\ttP} = \sum\mbox{}_{_{\mbox{\scriptsize I = 1}}}^{\sN} \,\underline{p}_I 
\mbox{ } , \mbox{ } \mbox{ } 
\underline{\ttL} = \sum\mbox{}_{_{\mbox{\scriptsize I = 1}}}^{\sN} \,\underline{q}^I\cr\underline{p}_I 
\mbox{ } , 
\eeq
so that the total momentum and the total angular momentum of the model universe are zero.

It is very straightforward \cite{06I} to pass from this to ERPM in terms of mass-weighted relative 
Jacobi coordinates. 
Doing so does not change the structure (c.f. below discussion of `Jacobi map'),
so that I often take the form in Jacobi coordinates as a starting point.  
Now,  
\beq
T = \sum\mbox{}_{_{\mbox{\scriptsize i = 1}}}^{\sn}||\Circ_{{\sB}}\underline{\iota}^i||^2/2 
\mbox{ } , \mbox{ } \mbox{ } 
\Circ_{{\sB}}\iota^{\alpha i} \equiv \dot{\underline{R}}^{i} - \dot{\underline{\mB}} \cr 
\underline{\iota}^{i} 
\eeq
(arbitrary Rot(N, $d$)-corrected velocities).  
The conjugate momenta are then 
\beq
\underline{\pi}_i = \delta_{ij}\Star_{\sB}\underline{\iota}^{j} \mbox{ } , \mbox{ } \mbox{ }
\Star_{\sB}\underline{R}^{i} \equiv 
\Star{\underline{R}}^{i} - \Star{\underline{\mB}} \cr \underline{R}^{i} \mbox{ } . 
\eeq
These obey as a primary constraint, quadratic and not linear in the momenta, the `energy constraint' 
\beq
\ttQ \equiv \sum\mbox{}_{_{\mbox{\scriptsize i = 1}}}^{\sn}\, \pi_i\mbox{}^2/2  + V = E \mbox{ } . 
\label{EC}
\eeq
The left hand side of this is also the restricted Hamiltonian for the system. 
Variation with respect to $\underline{\mB}$ gives as a secondary constraint linear in the momenta 
\beq
\underline{\ttL} = \sum\mbox{}_{_{\mbox{\scriptsize i = 1}}}^{\sn}\,\underline{\iota}^i 
\cr \underline{\pi}_i   \mbox{ } , \mbox{ } \mbox{ } 
\eeq
so that the total angular momentum of the model universe is zero (that the total momentum is 0 is 
already encoded in the use of a set of relative variables such as the $\underline{\iota}^i$).  
Thus one can see there is a $q^I \longrightarrow R^i$ `Jacobi map' which preserves the form of a large 
number of features of the theory (sending $I$-indexed objects to sets of 1 less $i$-indexed objects but 
with the same structure and with $\mu_i$ in place of $m_I$).
I.e., the moment of inertia, kinetic energy, dilational object and total angular momentum in relative 
Jacobi coordinates look just like their particle position counterparts in this sense.

\subsection{Cartesian and Dragt maps' part-preservation of structure}

In furthermore passing from spherical form to surrounding flat space $u^{\Delta}$ form, there is a 
trivial Cartesian map for N-stop metroland: $u^{\Delta} \longrightarrow n^{i}$ for these the Cartesian 
directions of the surrounding space.  
However, for triangleland, one has instead the rather less trivial Dragt map (\ref{dragt1}, \ref{dragt2}, 
\ref{dragt3}).  
Note that the Dragt map does preserve the form of a number of objects, though it is not quite as nice as 
the Jacobi map in this way:  $|\sum_{\Delta = 1}^3\mbox{Dra}^{\Delta}|^2 = I^2 = |\sum_{i = 1}^2 I^i|^2$, 
$|\sum_{\Delta = 1}^3\Pi_{\sD\sr\sa\, \Delta}|^2 = 2T = |\sum_{i = 1}^2\Pi_i|^2$ 
%
%
$\sum_{\Delta = 1}^3\underline{\mbox{Dra}}^{\Delta} \cdot \underline{\Pi}_{\sD\sr\sa\, \Delta} = 
2\ttD = 2\sum_{i = 1}^2 \underline{\iota}^i \cdot \underline{\pi}_{i}$.  
However, $ \sum_{\Delta = 1}^{3}\underline{\mbox{Dra}}^{\Delta} \cr 
\underline{\Pi}_{\sD\sr\sa\, \Delta}$ is nothing like $\sum_{i = 1}^2 \underline{\iota}^i \cr 
\underline{\pi}_i$.

\subsection{Direct Relationalspace implementation of configurational relationalism}

Given a Riemannian geometry $\langle\sigma$, $\bg\rangle$ [for $\bg$ a metric over a topological space 
$\sigma$; Riemannian suffices in the present paper's context], the natural mechanics in the sense of 
Jacobi and of Synge \cite{Lanczos} associated with it is 
\beq
\fI = \sqrt{2}\int \d\lambda\sqrt{{\cal K}_{\sr\se\sll}{\cal W}} 
\mbox{ } , \mbox{ } \mbox{ where } \mbox{ }
{\cal K}_{\sr\se\sll} = ||\dot{\mbox{\boldmath${\cal Q}$}}||_{\sbg}\mbox{}^2 \mbox{ } 
\eeq
is twice the kinetic term, ${\cal Q}$ are generalized configuration space coordinates and $\bg$ is the 
configuration space metric and `rel' stands for `relationalspace approach'.

The shape space of N-stop metroland is $\langle\mathbb{S}^{\sn - 1}$, $\bg_{\sss\sp\sh\se}$: the 
standard spherical metric$\rangle$, so that the natural SRPM associated with this is
\beq
\fI =  \sqrt{2}\int \d\lambda\sqrt{\fK_{\sr\se\sll}\fW} 
\label{genac}
\eeq
for 
\beq
\fK^{\sN-\sss\st\so\sp\,\sS\sR\sP\sM}_{\sr\se\sll} = 
||\dot{\mbox{\boldmath${\Theta}$}}||_{\sbg_{\ts\tp\th\te}}\mbox{}^2 = 
\sum\mbox{}_{\mbox{}_{\mbox{\scriptsize $\barr = 1$}}}^{\sn - 1}
\prod\mbox{}_{\mbox{}_{\mbox{\scriptsize $\hat{p} = 1$}}}^{\barr - 1}
\mbox{sin}^2\Theta_{\hat{p} }\dot{\Theta}_{\barr}^2 \mbox{ } , 
\eeq 
cast in terms of ultraspherical coordinates.

The shape space of N-a-gonland is $\langle\mathbb{CP}^{\sn - 1}$, $\bg_{\sF\sS}$: the Fubini--Study 
metric$\rangle$, so that the natural SRPM associated with this is (\ref{genac}) with, in inhomogeneous 
coordinates 
\beq
\fK^{\sN-\sa-\sg\so\sn\,\sS\sR\sP\sM}_{\sr\se\sll} = 
||\dot{\mbox{\boldmath${\cal Z}$}}||_{\sbg_{\tF\tS}}\mbox{}^2 = 
\{   { \{1 + ||\mbox{\boldmath${\cal Z}$}||_{\sc}^2\} 
                ||\dot{\mbox{\boldmath${\cal Z}$}}||_{\sc}^2 - 
                 |(\mbox{\boldmath${\cal Z}$} \cdot \dot{\mbox{\boldmath${\cal Z}$}})_{\sc}|^2    }/
            {    \{1 + ||\mbox{\boldmath${\cal Z}$}||_{\sc}^2\}^2    } \} \mbox{ } .
\eeq
In the case of triangleland, this simplifies to 
\beq
\fK^{\triangle\,\sS\sR\sP\sM}_{\sr\se\sll} = 
|\dot{\cal Z}|^2/\{1 + |{\cal Z}|^2\}      = 
\{\dot{\cal R}^2 + {\cal R}^2\dot{\Theta}^2\}/\{1 + {\cal R}^2\}^2
\label{Jup}
\eeq
in polar coordinate form.
This triangleland case is also then castable in terms of $\langle\mathbb{S}^2$, 
$\bg_{\sss\sp\sh\se}$(radius 1/2)$\rangle$, so that e.g. (passing to a barred banal representation that 
absorbs the $(1/2)^2$ factor into the potential): 
\beq
\fI =  \sqrt{2}\int \d\lambda\sqrt{\bar{\fK}_{\sr\se\sll}\bar{\fW}} 
\mbox{ }\mbox{ with } \mbox{ }
\bar{\fK}^{\triangle\, \sS\sR\sP\sM}_{\sr\se\sll} = \dot{\Theta}^2 + \mbox{sin}^2\Theta\,\dot{\Phi}^2 \mbox{ } ,
\label{slik}
\eeq
in spherical coordinates ($\Theta, \Phi$) [ = ($\Theta_1, \Theta_2$)].

The relational space of N-stop metroland is C($\mathbb{S}^{\sn - 1}$, $\bg_{\sss\sp\sh\se}$) = 
$\langle\mathbb{R}^{\sn}$, $\bg_{\sf\sll\sa\st}\rangle$, so that the natural ERPM associated with this 
is
\beq
\fI =  \sqrt{2}\int \d\lambda\sqrt{K_{\sr\se\sll}W} 
\mbox{ } \mbox{ with } \mbox{ }
K^{\sN-\sss\st\so\sp\,\sE\sR\sP\sM}_{\sr\se\sll} = ||\dot{\mbox{\boldmath${\iota}$}}||\mbox{}^2 = 
\dot{\iota}^2 + 
\iota^2\sum\mbox{}_{\mbox{}_{\mbox{\scriptsize $\barr = 1$}}}^{\sn - 1}
\prod\mbox{}_{\mbox{}_{\mbox{\scriptsize $\hat{p} = 1$}}}^{\barr - 1}
\mbox{sin}^2\Theta_{\hat{p} }\dot{\Theta}_{\barr}^2
\label{uultra}
\eeq
in the shape-scale split's ultraspherical polar coordinates.

The relational space of N-a-gonland is C($\langle\mathbb{CP}^{\sn - 1}$, $\bg_{\sF\sS}\rangle$) = 
$\langle\mC(\mathbb{CP})^{\sn - 1}$, $\bg_{\sC(\sF\sS)}\rangle$, so that the natural ERPM associated 
with this is (\ref{slik}), 
\beq
K^{\sN-\sa-\sg\so\sn\,\sE\sR\sP\sM}_{\sr\se\sll} = \dot{\iota}^2 + 
\iota^2||\dot{\mbox{\boldmath${\cal Z}$}}||_{\sbg_{\tF\tS}}\mbox{}^2 
= \dot{\iota}^2  + \iota^2
\{{    \{1 + ||\mbox{\boldmath${\cal Z}$}||_{\sc}^2\} 
                ||\dot{\mbox{\boldmath${\cal Z}$}}||_{\sc}^2 - 
                 |(\mbox{\boldmath${\cal Z}$} , \dot{\mbox{\boldmath${\cal Z}$}})_{\sc}^2    }\}/
            {    \{1 + ||\mbox{\boldmath${\cal Z}$}||_{\sc}^2\}^2    } \mbox{ } ,
\label{Cripps}
\eeq
in $\iota$ alongside inhomogeneous coordinates ${\cal Z}^{\bar{r}}$. 
In the case of triangleland, this simplifies to 
\beq
K^{\triangle\,\sE\sR\sP\sM}_{\sr\se\sll} = 
\{\dot{\iota^2} + \iota^2|\dot{{\cal Z}}|^2\}/\{1 + |{\cal Z}|^2\} = 
\dot{\iota^2} + \iota^2\{\dot{\cal R}^2 + {\cal R}^2\dot{\Theta}^2\}/\{1 + {\cal R}^2\}^2 \mbox{ }   
\label{Hermes}
\eeq
in polar coordinate form.  
This triangleland case is also then castable in terms of $\langle\mathbb{S}^2$, 
$\bg_{\sss\sp\sh\se}$(radius 1/2)$\rangle$, so that e.g. 
\beq
\fI =  \sqrt{2}\int \d\lambda\sqrt{\bar{K}_{\sr\se\sll}\bar{W}} \mbox{ } ,
\mbox{ } \mbox{ with } \mbox{ }
K^{\triangle\,\sE\sR\sP\sM}_{\sr\se\sll} = 
\dot{\iota^2} + \{\iota^2/4\}\{\dot{\Theta}^2 + \mbox{sin}^2\Theta\,\dot{\Phi}^2\} \mbox{ } 
\label{Aphrodite}
\eeq
in spherical coordinates.
Then use $I$ instead to obtain a conformally flat metric form, 
\beq
K^{\triangle \,\sE\sR\sP\sM}_{\sr\se\sll} = 
\{1/4I\}\{\dot{I^2} + I^2\{\dot{\Theta}^2 + \mbox{sin}^2\Theta\,\dot{\Phi}^2\}\} \mbox{ } ,
\label{Apollo}
\eeq
and pass to a new banal representation (conformal factor $4I$: the checked one) in which that is the 
flat metric gives 
\beq
\check{K}^{\triangle \,\sE\sR\sP\sM}_{\sr\se\sll} = 
\dot{I^2} + I^2\{\dot{\Theta}^2 + \mbox{sin}^2\Theta\,\dot{\Phi}^2\} \mbox{ }  
\eeq
(away from $I = 0$ in which place this conformal transformation is invalid).  
\beq 
\fI =  \sqrt{2}\int \d\lambda\sqrt{\check{K}_{\sr\se\sll}\check{W}}
\label{genac4}
\eeq
and $\check{W} = W/4I$
Cartesianizing that, one ends up in Dragt coordinates, $K^{\triangle \sE\sR\sP\sM}_{\sr\se\sll} = 
\sum\mbox{}_{\mbox{}_{\Gamma = 1}}^3\{\dot{\mbox{Dra}}^{\Gamma}\}^2$.

\subsection{Reduction approach}\label{SSec: Red}

Use mass-weighted Jacobi coordinates, alongside $I = ||\mbox{\boldmath${\iota}$}||^2$:   
$K = ||\dot{\mbox{\boldmath${\iota}$}}||^2$, $\fK = IK$, and the stationary frame version of $\ttD$, 

\noindent
$D \equiv (\mbox{\boldmath${\iota}$} \cdot \dot{\mbox{\boldmath${\iota}$}})$.  
Then $\uttL$ and $\ttD$ give, in Lagrangian form, 

\noindent
\beq
\underline{\ttL} = \sumin\uiota^{i}\cr\{\dot{\uiota}^{i} - \dot{\underline{\mB}} \cr \uiota^{i} + 
\dot{\mC}\uiota^{i}\} = 0 
\mbox{ } , \mbox{ } 
\ttD = \sumin\uiota^{i}\cdot\{\dot{\uiota}^{i} - \dot{\underline{\mB}} \cr \uiota^{i} + 
\dot{\mC}\uiota^{i}\} = 0 
\mbox{ } .
\eeq
Now the third term of the first equation and the second term of the second equation are 0 by 
symmetry-antisymmetry, so eliminating (`Routhian reduction', see e.g. \cite{Lanczos, Goldstein}) 
$\dot{\underline{\mB}}$ from the first and $\dot{\mC}$ from the second in no way interfere with each 
other. 
The first equation then gives, $\sum_{i = 1}^{n}\underline{\uiota}\cr\{\uiota\cr\dot{\underline{\mB}}\} = 
-\sum_{i = 1}^n\uiota^i\cr\dot{\uiota}^i$, which is recastable, at least formally, as 
$\dot{\underline{\mB}} = -\underline{\underline{I}}^{-1}\underline{L}$ for $\underline{L}$ the 
stationary frame version of $\uttL$ and $\underline{\underline{I}}$ the barycentric inertia tensor,  
\beq
{I}_{\alpha\beta} = \sumin 
\left\{
|\iota^i|^2\delta_{\alpha\beta} - \iota^i_{\alpha}\iota^i_{\beta}  
\right\}
\mbox{ }  .  
\eeq
This is realizable in 2-$d$ away from the cone-point $I = 0$, but has further singularities in 3-$d$ on 
the collinear configurations (due to these having a zero principal moment factor for collinearities in 
the 3-$d$ case, and these not being cases one has any particular desire to exclude on physical grounds).  
The second equation gives $\dot{\nC} = -I^{-1}D$ (realizable away from $I = 0$, which is never in any 
case included in SRPM).

Then 
\beq
\fK^{\sS\sR\sP\sM}_{\sr\se\sd} = \{I\{K - A\} - D^2\}/I^2 \mbox{ } 
\label{Centauri}
\eeq 
for $A = \underline{L}\underline{\underline{I}}\underline{L}$ (twice the rotational kinetic energy).   
So, for 1-$d$, as $\underline{L} = 0$, $A = 0$ and the expressions for $I$, $K$ and $D$ give this to be 
twice the ultraspherical kinetic term in Beltrami coordinates, 
\beq
\fK^{\sN-\sss\st\so\sp\,\sS\sR\sP\sM}_{\sr\se\sd} = 
\{||\mbox{\boldmath${\iota}$}||^2||\dot{\mbox{\boldmath${\iota}$}}||^2 - 
(\mbox{\boldmath${\iota}$}\cdot \dot{\mbox{\boldmath${\iota}$}})^2\}/
\{||\mbox{\boldmath${\iota}$}||^2\}^2 \mbox{ } .  
\eeq
Then recasting this in terms of ultraspherical coordinates,  
$\fK^{\sN-\sss\st\so\sp\,\sS\sR\sP\sM}_{\sr\se\sd}$ is identified to be the same as the 
$\fK^{\sN-\sss\st\so\sp\,\sS\sR\sP\sM}_{\sr\se\sll}$ of (\ref{uultra}).  
%
%
For 2-$d$, A has another form by the inertia tensor collapsing to just a scalar, the definition of 
$\underline{L}$ 
and the Kronecker delta theorem, 
$A = I^{-1}\sum_i\sum_j\{(\uiota^{i}\cdot\uiota^j)(\dot{\uiota}^{i}\cdot\dot{\uiota}^j) - 
                         (\uiota^{i}\cdot\dot{\uiota}^j)(\uiota^{j}\cdot\dot{\uiota}^i)\}$. 
Then using multipolar Jacobi coordinates, 
\beq
A = I^{-1}\sum\mbox{}_{\mbox{}_{\mbox{\scriptsize i = 1}}}^{\sn} 
          \sum\mbox{}_{\mbox{}_{\mbox{\scriptsize j = 1}}}^{\sn}
          \{\iota^i\}^2\{\iota^j\}^2\dot{\theta}^i\dot{\theta}^j 
\mbox{ } \mbox{ and } \mbox{ }
D^2 =     \sum\mbox{}_{\mbox{}_{\mbox{\scriptsize i = 1}}}^{\sn}
          \sum\mbox{}_{\mbox{}_{\mbox{\scriptsize j = 1}}}^{\sn}
          \iota^i\dot{\iota}^i\iota^j\dot{\iota}^j \mbox{ } , 
\eeq
so 
\beq
IA + D^2 = |(\bar{\bz}\cdot\bz)_{\sc}|^2
\eeq
in complex notation.  
Then as also $I = ||\mbox{\boldmath${\iota}$}||^2 = ||\bz||_{\sc}\mbox{}^2$ and 
$K = ||\dot{\mbox{\boldmath${\iota}$}}||^2 = ||\dot{\bz}||_{\sc}\mbox{}^2$, one obtains 
twice the Fubini--Study kinetic term in inhomogeneous coordinates, thus identifying  
$\fK^{\sN-\sa-\sg\so\sn\,\sS\sR\sP\sM}_{\sr\se\sd}$ to be the same as Sec 3.4's  
$\fK^{\sN-\sa-\sg\so\sn\,\sS\sR\sP\sM}_{\sr\se\sll}$ corresponding to Sec 2.6's line element 
$\d s^2_{\sN-\sa-\sg\so\sn\,\sS\sR\sP\sM}$.    
The triangleland case is then simpler as per (\ref{Jup}) and rearrangeable to (\ref{slik}) by the usual 
moves.

Next, since there is now no dilational constraint, and in the mechanical rather than geometrical banal 
representation, 
\beq
K^{\sE\sR\sP\sM}_{\sr\se\sd} = K - A = \{D/\sqrt{I}\}^2 + I\fK^{\sS\sR\sP\sM}_{\sr\se\sd}  \mbox{ } .
\eeq
[The second equality is by (\ref{Centauri}).].  
Then $D = (\mbox{\boldmath${\iota}$}\cdot\dot{\mbox{\boldmath${\iota}$}}) = 
\dot{I}/2 = \iota\dot{\iota}$, so $D/\sqrt{I} = \dot{\iota}$, and so 
\beq
K^{\sE\sR\sP\sM}_{\sr\se\sd} = \dot{\iota}^2 + \iota^2\fK^{\sS\sR\sP\sM}_{\sr\se\sd} \mbox{ } , 
\eeq
i.e. twice the reduced ERPM kinetic term is the one whose metric is the cone over the metric corresponding to 
twice the reduced SRPM kinetic term.  
Note furthermore that this derivation is independent of the spatial dimension.
Next, in the cases for which $\fK^{\sS\sR\sP\sM}_{\sr\se\sd}$ has been specifically derived, 
$K^{\sE\sR\sP\sM}_{\sr\se\sd}$ can then also immediately be derived.  
Namely, in 1-$d$, $\fK^{\sN-\sss\st\so\sp\,\sE\sR\sP\sM}_{\sr\se\sd}$ is identified to be the same as 
the $\fK^{\sN-\sss\st\so\sp\,\sE\sR\sP\sM}_{\sr\se\sll}$ of (\ref{uultra}).  
Of course, this case could have been established more trivially, since in this case there are no 
constraints to eliminate, leaving this working being but a change to the shape-scale-split-abiding 
ultraspherical coordinates.
However, the 2-$d$ case is less trivial, amounting to $\fK^{\sN-\sss\st\so\sp\,\sS\sR\sP\sM}_{\sr\se\sd}$ 
being identified to be the same as the $\fK^{\sN-\sss\st\so\sp\,\sS\sR\sP\sM}_{\sr\se\sll}$ of 
(\ref{Cripps}), with the simpler triangleland example (\ref{Hermes}) rearrangeabale to the forms 
(\ref{Hermes}, \ref{Aphrodite}, \ref{Apollo}).

\subsection{Discussion of relationalspace approach versus reduced approach and NM split}

Thus (Theorem): in 1- and 2-$d$, the direct relationalspace implementation B) of Sec 3.4 is equivalent 
to the Barbour- and Barbour--Bertotti-type indirect implementation A) of Sec 3.2. 

\noindent 
This extends to the oriented rather than plain case as well, due to this just changing coordinate ranges, 
which did not form part of the above argument.  

\noindent Note that the reduction approach amounts to the counterpart of GR's thin-sandwich prescription being 
achievable for 1- and 2-$d$ RPM's.

\noindent Also, the above reduction is a reduction at the first of the following different possible 
levels.

\noindent C)I) Configuration space reduction.

\noindent C)II) Phase space reduction (e.g \cite{MarsWein}).  

\noindent C)III) Reduction at the level of the QM equations (e.g. in \cite{06I, RedDir}).  

\noindent [Reduction {\sl after} quantization is the {\sl Dirac quantization scheme}, though this 
usually involves reduction at a {\sl fourth} level 

\noindent C)IV) -- that of the QM solutions themselves; III) and IV) are further discussed in 
\cite{RedDir}.]

\noindent
D) Comparison with the Relational-Absolute split of Newtonian Mechanics. 

\mbox{ }

\noindent
Note i) D) is well-studied and provides coordinate systems that are useful in relational and reduction 
approaches as well:  Jacobi coordinates, spherical-type coordinates, Dragt coordinates and their 
extension to more than 3 particles, democratic invariants, studies of the somewhat harder oriented and 
3-$d$ cases, and other work of Iwai \cite{Iwai}, Montgomery \cite{Mont96, Mont98}  and Hsiang 
\cite{Hsiang}, and in work reviewed by Littlejohn and Reinsch \cite{LR}.

\noindent Note ii) Butterfield \cite{Butter} terms C)I) relational and C)II) reductive [Belot \cite{Be} 
also compares C)I) and C)II)].  
However, I consider these both to be reductions in different formalisms and coincident for many purposes 
[II) being set up more generally \cite{MarsWein} thus making it available for a wider range of 
examples], while I compare all of A) to D) including the four variants I) to IV) of C)].  
I find that (Sec 5, \cite{RedDir}) each of II, III and IV can differ due to interference of 
operator-ordering issues, and D) can likewise be distinct from A) to C).

\noindent Note iii) In D) the new relational coordinates are part of a coordinate system linked to the 
original coordinatess by sequence of coordinate trandformations, while B),C) involve new variational 
starting points, including taking the reduced action itself as a new starting point within scheme C).  

\noindent 
Note iv) The indirect formulation A) retains the virtue of being a gateway to GR analogy of use in 
Quantum Cosmology and the Problem of Time.  Thus recasting what can be worked out by B) or C) in terms 
of A) remains of interest.

\subsection{Equations of motion for the relational space or reduced formulation}

Next I consider the action in relationalspace first principles or reduced form for 3- and 4-stop metrolands and 
triangleland. 
For scaled 3-stop metroland, the action can be written as (\ref{prod}) with kinetic term in plane polar 
coordinates ($\iota$, $\varphi$).
The conjugate momenta are then 
\beq
p_{\iota} = \iota^* \mbox{ } , \mbox{ } \mbox{ } p_{\varphi} = \iota^2\varphi^* \mbox{ } .
\eeq
These obey as a primary constraint, quadratic and not linear in the momenta, the `energy constraint' 
\beq
\ttQ = p_{\iota}^2/2 + p_{\varphi}^2/{2\iota^2} + V  = E \mbox{ } ,
\eeq
the left hand side of which is also now the Hamiltonian for the system.  
The Euler--Lagrange equations of motion are then 
\beq
\iota^{**} - \iota{\varphi^*}^2 = -\pa V/\pa\iota \mbox{ } , \mbox{ } 
\{\iota^2\varphi^*\}^* = - \pa V/\pa\varphi \mbox{ } ,
\eeq
one of which can be replaced by the energy integral 
${\iota^*}^2/2 + \iota^2{\varphi^*}^2/2 = E - V$.

For scaled 4-stop metroland, the action is (\ref{prod}) with kinetic term in spherical polar coordinates 
($\iota$, $\theta$, $\phi$).  
Then the conjugate momenta are 
\beq
p_{\iota}  = \iota^*                               \mbox{ } , \mbox{ } \mbox{ } 
p_{\theta} = \iota^2\theta^*                       \mbox{ } , \mbox{ } \mbox{ } 
p_{\phi}   = \iota^2\mbox\,\mbox{sin}^2\theta\,\phi^*   \mbox{ } .
\eeq
These obey as a primary constraint, quadratic and not linear in the momenta, the `energy constraint' 
\beq
\ttQ =  p_{\iota}^2/2 + p_{\theta}^2/{2\iota^2} + p_{\phi}^2/{2\iota^2\mbox{sin}^2\theta} + V  = E 
\mbox{ } ,
\eeq
the left hand side of which is also the Hamiltonian for the system.  
The Euler--Lagrange equations of motion are then 
\beq
\iota^{**} - \iota\{\theta^{*\,2} + \mbox{sin}^2\theta\,\phi^{*\,2}\} = -\pa V/\pa\iota 
\mbox{ } , \mbox{ } 
\{\iota^2\mbox{sin}^2\theta\,\phi^*\}^* = -\pa V/\pa\phi 
\mbox{ } , \mbox{ } 
\{\iota^2\theta^*\}^* - 2\iota^2\mbox{sin}\theta\,\mbox{cos}\,\theta\,{\phi^*}^2 = -\pa V/\pa\theta 
\mbox{ } ,
\eeq
one of which can be replaced by the energy integral, 
${\iota^*}^2/2 + \iota^2\{\theta^{*\,2} + \mbox{sin}^2\theta\,\phi^{*\,2}\}/2 = E - V$.  
This form generalizes straightforwardly to the N-stop metroland case.

For scaled triangleland \cite{08I}, the action is (\ref{prod}) but in checked banal representation. 
One's equations of motion are then the same as for 4-stop metroland under ($\iota$, $\theta$, $\phi$, 
$E$, $V$) $\longrightarrow$ ($I$, $\Theta$, $\Phi$, $\check{E} = E/4I$, $\check{V}$), except that, as 
$\check{E}$ is now a function of $I$, the $I$ Euler--Lagrange equation picks up a new term, 
$I^{\check{*} \check{*}} - I\{\Theta^{\check{*}\,2} + \mbox{sin}^2\Theta\,\Phi^{\check{*}\,2}\} + 
{\pa V}/{\pa I} + E/4I^2 = 0$.

\subsection{Physical Interpretation of RPMs' conserved quantities}

Unoriented 3-stop metroland has isometry group $SO$(2), to which there corresponds an object ${\cal D} = 
\iota^2\varphi^*$, which is a constant of the motion in the case of $\varphi$-independent potentials.  
Physically, this is straightforwardly the relative dilational momentum of one of the clustering's 
subsystems relative to the other.
This model has a tight mathematical analogy with the central force problem in the plane whose sole  
conserved quantity is the angular momentum, $L$.

Spherical coordinates ($\alpha, \chi$) are good for solving many aspects of the dynamics of the 
sphere no matter what the physical interpretation is to be.  
The sphere's isometry group is $SO$(3).  
Denote the 3 $SO$(3) objects by ${\cal R}_{\Delta}$.
${\cal R}_{\sT\so\st\sa\sll} = \sum_{\Delta = 1}^3{{\cal R}}_{\Delta}\mbox{}^2$
One arises as a conserved quantity for $\chi$-independent potentials so $r^2\chi^*$  is constant, while 
${\cal R}_{\sT\so\st\sa\sll}$ is conserved if the potential is additionally $\alpha$-independent, i.e. a 
function of $r$ alone, which in the usual spatial case is termed `central'.  
Now, these ${\cal R}_{\Delta}$ are in general physically a somewhat unfamiliar generalization of angular 
momenta \cite{Smith} envisaged in \cite{AF} as `rational momenta' [i.e. corresponding to (functions 
of) ratios, of which angles are but a subcase, so `rational momenta' are physically a generalization of 
angular momentum; they are not however a mathematical generalization -- they still correspond to the 
same sort of Lie group mathematics that angular momentum corresponds to].  
These cover the following cases a) for the sphere in space these are angular momenta $L_i$ 
and $L_{\sT\so\st\sa\sll}$ and the usual quantum numbers m and l.  
b) For unoriented 4-stop metroland they are physically not angular momenta but rather relative dilational momenta 
${\cal D}_a$ and ${\cal D}_{\sT\so\st\sa\sll}$ with `projected'\footnote{This is termed 
`projected' in \cite{AF} to remove the rotation and atomic physics prejudices in the more common names 
`axial' and `magnetic'.} 
and total relative dilation quantum numbers d and D.  
c) For unoriented triangleland, they are a relative angular momentum $\tt{R}_3 = {\cal J}_3$ and 
two mixed relative angular momentum-- dilational momentum quantities ${\cal R}_1$ and 
${\cal R}_2$ and ${{\cal R}}_{\st\so\st\sa\sll}$ and projected and total quantum numbers j and R.   
4-stop metroland and triangleland are already covered in \cite{AF, +tri}, given \cite{+tri}'s correction 
of \cite{08I}'s incorrect claim that SRPM and ERPM conserved quantities are not quite the same.  
They are.  
How these results extend to quadrilateralland is also of interest \cite{QCons}.

\section{52 parallels between RPM's and GR-as-geometrodynamics}\label{SSec: para}

This is my main motivation for studying RPM's. 
First I lay out the GR counterparts of SSecs 3.1-2.  

\noindent
1) In GR-as-geometrodynamics, the role of $\fR$(n, $d$)'s is played by the space Riem($\Sigma$) of 
Riemannian 3-metrics on a fixed spatial topology $\Sigma$. 

\noindent 
2) The role of Rot(n, $d)$'s as group of irrelevant transformations $G$ is played by the  
3-$d$iffeomorphism group, Diff($\Sigma$).  
This analogy goes further through both groups being nonabelian (though, as that requires $d > 2$, this 
further aspect is beyond the scope of this paper's specific examples).  

\noindent 
3) The role of Dil as a further contribution to the group of irrelevant transformations is played in a 
certain sense by the conformal transformations Conf($\Sigma$) and in a certain sense by the 
volume-preserving conformal transformations, VPConf($\Sigma$) \cite{ABFKO}.  
This analogy goes further in the sense that these are all scales, though Dil is global (here in 
the sense of pertaining to the whole system rather than to any particular cluster therein), 
Conf($\Sigma$) is local and VPConf($\Sigma$) is `local excluding one global degree of freedom' (the 
global volume).  

\noindent 
4) Both for RPM's and for geometrodynamics, composing 2) and 3) involves a semidirect product \cite{I84}.

\noindent 
5) Relational space = $\fR(\mN, d)$/Rot($d$) corresponds to superspace($\Sigma$) = 
Riem($\Sigma$)/Diff($\Sigma$) \cite{DeWitt67, DeWitt70, Fischer}.  

\noindent 
6) Preshape space is the analogue of a little studied space ``CRiem($\Sigma$)" \cite{08I} alias 
``pointwise version of CS($\Sigma$)" \cite{FM96}.  

\noindent
7) Shape space = $\fR(\mN, d)$/Rot $\times$ Dil for is analogous to conformal superspace 
\cite{York74, ABFO} CS($\Sigma$) = Riem($\Sigma$)/Diff($\Sigma$) $\times$ Conf($\Sigma$) \cite{York7273, 
York72Time, York74, ABFO, FM96}.

\noindent 
8) Both RPM's and GR admit shape-scale splits.  
For RPM's, the role of scale is played by such as $\sqrt{I}$ or $I$ or $\half\mbox{ln}\,I$ = ln\,$\iota$, 
and for GR it is played by such as the scalefactor $a$, 3-metric determinant (alias local volume) 
$\sqrt{h}$, conformal factor $\phi$ or Misner scale variable $\Omega$ [$\sqrt{h} = a^6$, 
$\phi = \sqrt{a}$, $\Omega = -\mbox{ln}(a)$].  

\noindent 9) The GR counterpart of relationalspace being the cone over shape space in RPM's is the 
configuration space \{CS + V\}($\Sigma$) = Riem($\Sigma$)/Diff($\Sigma$) 
$\times$ VPConf($\Sigma$), where the V stands for global volume. 
This is the closest thing known to a `space of true dynamical degrees of freedom for GR' \cite{York7273, 
ABFKO}.
While there are indications that (structures conformal to) cones also play a role in the study 
of GR configuration spaces, I leave this issue for a future occasion.  
Another manifestation of the analogy is that each of relational space and superspace \cite{K81} has a 
conformal Killing vector associated with scale.  
 
\noindent 
10) Additionally, these configuration spaces are in general stratified for both GR and RPM's 
(see \cite{DeWitt70, Fischer} and Sec 2.3).  

\noindent 
11) Also, for both GR and RPM's, many of the configuration spaces have physically-significant bad 
points (e.g. $a = 0$ is the Big Bang and $I = 0$ is the maximal collision).

\noindent 12) The tessellation by geometrical/physical interpretation method of SSec 2.7 has a 
counterpart that is also useable in simple minisuperspace examples to read off the meaning of classical 
dynamical trajectories or of the QM wavefunction. 

\mbox{ } 

\noindent Analogies between RPM's and GR at the level of the action are as follows.  

\noindent 13) There is a relational action \cite{RWR} for GR-as-geometrodynamics\footnote{The spatial topology 
$\Sigma$ is taken to be compact without boundary. 
$h_{\mu\nu}$ is a spatial 3-metric thereupon, with determinant $h$, covariant derivative $D_{\mu}$, 
Ricci scalar Ric($h$) and conjugate momentum $\pi^{\mu\nu}$.  
$\Lambda$ is the cosmological constant.
$M^{\mu\nu\rho\sigma} = h^{\mu\rho}h^{\nu\sigma} - h^{\mu\nu}h^{\rho\sigma}$ is the inverse 
DeWitt supermetric with determinant $M$ and inverse $N_{\mu\nu\rho\sigma}$. 
To represent this as a configuration space metric (i.e. with just two indices, and 
downstairs), use DeWitt's 2 index to 1 index map \cite{DeWitt67}.
$\dot{\sF}^{\mu}$ is the velocity of the frame; in the manifestly relational formulation of GR, this 
cyclic velocity plays the role more usually played by the shift Lagrange multiplier coordinate.
$\pounds_{\dot{\tF}}$ is the Lie derivative with respect to $\dot{\sF}^{\mu}$.} 
\beq
\fI^{\sG\sR} = 
2\int\d\lambda\int\d^{3}x\sqrt{h}\sqrt{\fT^{\sG\sR}_{}
\{\mbox{Ric}(h) - 2\Lambda\}} \mbox{ } \mbox{  for } \mbox{ }   
\fT^{\sG\sR}_{} = 
\frac{1}{4}M^{\mu\nu\rho\sigma}\Circ_{\sF}h_{\mu\nu}\Circ_{\sF}h_{\rho\sigma} \mbox{ }, \mbox{ } \mbox{ }
\Circ_{\sF}h_{\mu\nu} = \dot{h}_{\mu\nu} - \pounds_{\dot{\sF}}h_{\mu\nu}
\label{GRaction} \mbox{ } 
\eeq
that is the immediate counterpart of the indirectly-formulated ERPM action.    
There is also a relational CS+V action \cite{ABFKO} that has further parallels with the shape-scale 
split form of the ERPM action (and the conformal gravity action has further parallels with SRPM: here 
homogeneity requirements are taken care of, in both cases by incorporation of powers of some 
dimensionful quantity: $I$ in SRPM and V in an alternative relational theory of conformal gravity 
\cite{ABFO}). 

\noindent 14) Reading off (\ref{prod}) and (\ref{GRaction}), energy $E$ and cosmological constant 
$\Lambda$ (up to factor of $- 2$) play an analogous role at the 

\noindent 
level of the relational actions, so to 
some extent $E$ is a toy model for the mysteries surrounding $\Lambda$. 
[As regards getting rid of $\Lambda$, Barbour showed that $E = 0$ for SRPM \cite{B03}, but then I 
showed (e.g. \cite{08I}) that an analogous pseudoenergy $\fE$ of physical dimension 
Energy/(Moment of Inertia) slips into the theory instead.]       
Also note that this analogy fails to hold at the level of the equations of motion [c.f. 24)].  

\noindent
The way in which that the physical equations follow from the relational action for GR-as-geometrodynamics 
and from the indirectly formulated RPM actions then have many parallels.

\noindent 15) Both subsequent workings involve their own version of an emergent time.  
The standard versions of these are the Jacobi--Barbour--Bertotti recovery \cite{BB82, B94I} of Newtonian 
time but on relational foundations, and an emergent time in GR that reduces to cosmic time under 
appropriate cosmological modelling circumstances.    

\noindent 16) By reparametrization invariance \cite{Dirac}, each has a primary constraint quadratic in 
the momenta, so that there is an analogy between the form of, and means of arriving at, the GR Hamiltonian constraint
\beq
{\cal H} \equiv {N}_{\mu\nu\rho\sigma}\pi^{\mu\nu}\pi^{\rho\sigma} - \sqrt{h}\{\mbox{Ric}(h) - 
2\Lambda\} = 0 
\label{HamCo}
\eeq
and the ERPM `energy constraint' (\ref{EC}).   
This is the basis of a number of further analogies between mechanics results and GR results [c.f. 36) and 
37)].  

\noindent 17) By variation with respect to the auxiliary $G$-variables, each relational theory has 
constraints linear in the momenta: for GR, the momentum constraint (21 ii) 
\beq
{\cal L}_{\mu} \equiv - 2D_{\nu}{\pi^{\nu}}_{\mu} = 0 \mbox{ } 
\label{GRmom}
\eeq
from variation with respect to $\mF^{\mu}$, and, for RPM's, the zero total angular momentum constraint 
$\ttL_{\mu} = \sumin\mbox{ } \{\uRR^i \cr \uPP_i\}_{\mu} $ 
(and SRPM's zero total dilational momentum constraint, $\ttD \equiv \sumin\mbox{ } \uRR^i \cdot \uPP_i 
= 0$). 
These arise from variation with respect to $\mB^{\mu}$ (and of a dilational auxiliary). 

\noindent 18) The zero total dilational momentum constraint $\ttD$ moreover closely parallels the well-known GR 
maximal slicing condition \cite{Lich, York7273}, $h_{\mu\nu}\pi^{\mu\nu} = 0$.  

\noindent 19) `Dilational' conjugates to scale quantities are e.g. the Euler quantity $\ttD$ \cite{06II} 
that is conjugate to $\mbox{ln}\,\iota$ and the York quantity \cite{York72Time} 
$Y = \frac{2}{3}h_{\mu\nu}\pi^{\mu\nu}/\sqrt{h}$ that is conjugate to $\sqrt{h}$. 
The York quantity is proportional to the mean curvature; it indexes constant mean curvature (CMC)  
foliations of (regions of) spacetime.  
This additional geometrical interpretation would appear to have no counterpart in RPM's.  
York's conformogeometrodynamical formulation (widely used in numerical relativity \cite{BS} as well 
as in the below discussion of internal time) is also closely tied to \{CS + V\}($\Sigma$).

\mbox{ } 

\noindent 
Some differences between RPM's and geometrodynamics are as follows. 
The constraint algebra for RPM's does not `crisscross' 
(i.e. the constraints are not integrability conditions for each other),  
unlike occurs in the Dirac Algebra of geometrodynamics \cite{Dirac, HKT}.   
Thus, for GR but not for RPM's, if one has only some of the constraints, one discovers the missing ones 
as integrabilities \cite{OMSan}.  
Also, RPM's also have nothing like the embeddability/hypersurface deformation interpretation of the 
Dirac Algebra (nor should there be as that is very diffeomorphism-specific) and thus nothing like the 
Hojman--Kucha\v{r}--Teitelboim \cite{HKT} first-principles route to geometrodynamics.  
Finally, there is no counterpart of the `relativity without relativity' \cite{RWR} first-principles 
route to geometrodynamics in the sense that the configuration space variables are not metrics (which 
leads to a tight restriction on possible potential-forming concomitants, unlike the great freedom one 
has in RPM's on the form of the potential.)
[One can however view the shape-scale split approach to ERPM as paralleling the first-principles route 
to conformogeometrodynamics 
in \cite{ABFKO}, as per 3), 4), 8), 12).]

\mbox{ } 

\noindent 20) The reduction of ERPM in Sec 3.5 is a successful counterpart of GR's thin sandwich 
conjecture \cite{BSW, ThinSan}.

\noindent 21) Scale in ERPM's is a direct counterpart of scale (and homogeneous matter modes in cases in 
which these are present) as heavy, slow H degrees of freedom, with shape versus inhomogeneous modes as 
light, fast L degrees of freedom (in GR, anisotropy could be an alternative L or part of a bigger H or part of a 
3-tier hierarchy: homogeneous isotropic H degrees of freedom, anisotropic `middling heaviness and 
middling slowness' `M' degrees of freedom and inhomogeneous L degrees of freedom).  

\noindent 22-26) are further analogies between particular RPM's and particular well-known simple early 
universe cosmology models which are presented in Sec 7.2.

\noindent 27)  At the level of kinematical quantization, RPM's are a subset of the toy model examples 
given in \cite{I84}; now that the groups involved can be decomposed as semisimple products [4)] is 
relevant as Mackey theory permits access to the representation theory in such cases.
However, the representation theory of the diffeomorphisms certainly has difficulties not present in the 
1 and 2-$d$ RPM's (for which the groups $SO$(N -- 1) and $SU$(N -- 1) occur, which are of course 
familiar from elementary representation theory and particle physics).  

\noindent 28)  There may be parallels between oriented choice of shapes and the affine approach to 
geometrodynamics \cite{I84}, 
as an extension of the half-line toy model for the affine case \cite{I84}.  

\noindent 29)  In each case, the presence of additional linear constraints permits one to choose to 
attempt the Dirac quantization approach.  
Geometrodynamics, not being reducible in general, that is one of few schemes that can be attempted in 
that case.  

\noindent 30)  One can give RPM's operator orderings in close parallel with those suggested for GR 
Quantum Cosmology (Laplacian, 

\noindent conformal and $\xi$ operator orderings, see Sec 5).  


\noindent 31) Whether the Guichardet connection \cite{Guichardet} and configuration space monopole 
issues of RPM's (see Sec 6 and Appendix) have counterparts for GR is an interesting question (addressable 
at least at a formal level).    

\noindent 32) Importantly, the structure shared by the Hamiltonian constraint of GR and the energy 
constraint for RPM's gives a frozen wave equation $\widehat{\cal H}\Psi = 0$, which is one well-known 
facet of the Problem of Time \cite{K92, I93}; this is the well-known Wheeler--DeWitt equation 
\cite{Battelle, DeWitt67}.  

\noindent 33) Both RPM's and geometrodynamics manifest the {\it Multiple Choice Problem} facet of the 
Problem of Time \cite{K92, I93}: the Groenewold--Van Hove phenomenon indeed already occurs for 
finite theories and thus can be expected to occur here just as it occurs in minisuperspace.    
Different choices of timefunction will in general give different quantum theories. 

\noindent 34) As regards the {\it Global Problem of Time} \cite{K92, I93}, I subdivide this into 
globality in time and in space. 
Then as regards globality in time, for GR, CMC slice existence and propagation and monotonicity does 
hold for a range of examples but by no means in all cases; the RPM analogue of this, involving equations 
along the lines of the well-known {\it Lagrange--Jacobi equation} 
of Celestial Mechanics, is likewise a guarantee in some but not all cases (see \cite{IT} for more 
details).  
These are only loosely analogous because the corresponding scale variables are not analogous, but both 
admit generalizations to directly analogous scale variables for which the analogy is tighter \cite{IT}.  

\mbox{ }

\noindent On the other hand, globality in space is a field-theoretic issue not expected to have a 
counterpart for RPM's, while the Torre impasse \cite{Torre} is specific to embedding variables.  
\noindent More generally, diffeomorphism-based approaches \cite{I84, IshKu} do not have much 
useful analogy in RPM's.
Further differences as regards other Problem of Time facets are as follows.

\noindent RPM's are `lucky' in Dirac's sense \cite{Dirac} and so have no {\it Functional Evolution 
Problem} \cite{K92, I93}. 
\noindent RPM's have nothing like the {\it Foliation Dependence Problem} \cite{K92, I93} as the 
embedding meaning of the GR's Dirac Algebra of constraints is lost through toy-modelling it with 
rotations (and/or dilations).  
\noindent  The RPM counterpart of the {\it Thin Sandwich Problem} is already covered in Sec 3; for RPM's 
this is not a problem but rather a resolved situation that opens up extra paths and checks.

\noindent As further differences, RPM's permit further approaches to quantization that are not open for 
(full or midisuperspace) GR: reduced quantization, and further operator ordering issues from whether one 
conformal/Laplacian/$\xi$ orders before or after reduction, and comparison with the relational--absolute 
split of Newtonian Mechanics (Sec 5, \cite{RedDir}).  
Nor do RPM's serve to model specifically infinite dimensional/field theory issues including whether the 
Wheeler--DeWitt equation is well-defined.  

\mbox{ }  

\noindent
35) In both cases the scale is cleanly split from the shape, so that it is tempting to use some scale 
variable as a time variable, e.g. scalefactor time, Misner time, local volume time in GR versus moment 
of inertia time, hyperradius time and ln\,$\iota$ time in RPM's, but in both theories this runs into 
monotonicity problems.   

\noindent 
36) In both cases, one can perform canonical transformation such that conjugate dilational objects are 
now the times, `York time' variable \cite{York72Time, IM, K81, K92, I93} $t^{\sY\so\sr\sk} \equiv Y =  
\frac{2}{3}h_{\mu\nu}\pi^{\mu\nu}/\sqrt{h}$, one can think of the passage from SRPM to ERPM as involving an analogous 
extra `Euler time' variable \cite{06II, SemiclI} $t^{\sE\su\sll\se\sr} \equiv \ttD = \sumin \uRR^i \cdot 
\uPP_i$.
This gives the aforementioned improved monotonicities by such as the Lagrange--Jacobi equation and the 
CMC condition's propagation 
equation.  
Also, SRPM corresponds to maximal slicing in being a case in which dilational time is frozen.  

\noindent
37) Then, for RPM's the equation to solve to isolate the new linearly-ocurring momentum $P_t$ (also 
then equal to minus the `true Hamiltonian') is a sum of powers of the scale variable, while, for GR, it 
is of the form $D^2\phi =$ sum of powers of $\phi$ (the {\it Lichnerowicz--York equation}). 
Note that by homogeneity the minisuperspace version of this is again just a rational polynomial.      
In each case, note that the powers involved can be of an exponential function so that the 
solution is a logarithm of a combination of roots and sums in soluble cases. 

\noindent 38) Changing scale variable nontrivially changes the equation that one is to solve by shifting 
what power of the scale the $t^2$ term comes with relative to the other rescalings of powers.  
As well as being relevant to RPM's, this may also be a useful move in studying minisuperspace \cite{IT}.  

\noindent 39) Both for RPM's and for minisuperspace, the hidden time working then becomes fraught with 
quantum-level operator ordering and well-definedness problems \cite{06I, SemiclI, IT}, and, moreover, with some mathematical 
differences between the two, stemming from e.g. definiteness versus indefiniteness and different 
manifestations of shape-scale split in each case, so that RPM's provide here extra examples of interest 
rather than just repeats of what minisuperspace already covers.
However, both for RPM's and for minisuperspace, one can apply the method of approximating in a series at 
the classical level and only then promoting the outcome of that to quantum operators \cite{IT}, 
which are then rather better defined and less ambiguous). 
This has parallels to the treatment of relativistic wave equations (done before in the semiclassical 
approach but not as far as I know for the hidden time approach).  

\noindent 40) There may be a `reference particles' analogue in RPM's of reference matter fields approach to 
hidden time.    

\noindent 41) Is there an analogue of $\Lambda$ providing a hidden time from e.g. analogy 13)  
or one of the analogies in 24)? 

\noindent 42) An emergent semiclassical time approach to the Problem of Time can be built on top of 
the aforementioned H-L split in each case. 
Here, the H-equation provides a $t^{\sW\sK\sB}$ aligned with $t^{\se\sm}$.
Then the L-equation becomes a $t^{\sW\sK\sB}$-dependent wave equation.
This scheme is discussed in more detail in Sec 9; it is analogous to Halliwell--Hawking's scheme 
\cite{HallHaw} for inhomogeneous perturbations about homogeneous 
semiclassical quantum GR, but now with a rather simpler coupled shape dynamics (approximately a time-dependent Schr\"{o}dinger equation). 
This extra simpleness is useful in investigating 

\noindent various features outlined in 47) below and in the 
Conclusion. 

\mbox{ }

\noindent Further non-analogies are as follows.  
Due to the difference in definiteness of the kinetic terms, the Schr\"{o}dinger inner product is 
available for RPM's but not GR, while the Klein--Gordon-type inner product (which fails for other 
reasons in GR \cite{K81}) is not available in RPM's.  
Also, the {\it third quantization} scheme makes no sense in RPM's, due to these being finite rather than 
field-theoretic.   

\mbox{ }

\noindent 43) RPM's admit toy models of the following timeless approaches.  
I) The {\it na\"{\i}ve Schr\"{o}dinger interpretation} \cite{HP86UW89}, in which 

\noindent simple questions of 
being are addressed, such as `what is the probability that the (model) universe is large? 
Approximately isotropic? 
Approximately homogeneous?' 4-stop metroland and triangleland toy model analogues of such questions are 
considered in \cite{AF, +tri}.
II) The {\it conditional probabilities interpretation} \cite{PW83}, in which questions concerning pairs of questions 
of being are addressed, such as `what is the probability that the (model) universe is large given 
that it is approximately isotropic?'  
In this case, furthermore, one of the two properties can have 
additional significance, e.g. it could be a good clock subsystem.

\noindent 44) RPM's are useful in the study of notion of locality in space (leading to notions of 
inhomogeneity and structure) and of locality in configuration space (leading to notions of states that 
are only approximately known), towards the timeless records approach to the Problem of Time.  
There are more options in RPM's, due to i) the kinetic term positive-definiteness lending itself to 
the construction of notions of locality in configuration space.  
ii) While in GR triviality of $D_{\mu}$ ties together ${\cal L}_{\mu}$ triviality and the lack of a notion of 
locality, in RPM's these notions are disjoint, the latter occurring even in the scaled 1-$d$ case 
that has no linear constraints at all.  

\noindent 45) There are also analogies useful for records theory at the level of notions of information, 
though these remain very much work in progress \cite{NOI}.   

\noindent 46) Histories theory schemes \cite{GMH, Hartle} can be set up for both RPM's and for GR.  

\noindent 47) ERPM, in particular of at least the quadrilateral, is well-suited for 
histories-records-semiclassical combination investigations, as a simple toy model, but nevertheless with 
sufficient features to do a reasonable job of toy-modelling midisuperspace.  
As explained in the Introduction, this combination is of particular interest, and a major goal of 
my work on RPM's. 

\mbox{ }  

\noindent Further foundational issues in Quantum Cosmology (possibly) addressable by RPM's are as follows.  

\noindent 48) RPM's are likely to be a useful example as regards the {\it Problem of Observables} (this 
is tied to evolving constants of the motion/perennials/partial observables approaches \cite{K92, I93, 
Rovelli} to the Problem of Time).  

\noindent 49) RPM's are a toy model \cite{+tri} for the role of {\it uniform states} in GR Quantum 
Cosmology; e.g. in triangleland, the equilateral triangle is the most uniform configuration \cite{+tri}.  

\noindent 50) RPM's are a toy model for closed universe issues \cite{DeWitt67}, RPM examples of which 
are energy interlocking (i.e. the energies of the various subsystems add up to a fixed total energy 
of the universe) and angular momentum counterbalancing (i.e. the angular momenta of the various 
subsystems add up to zero by $\uttL = 0$) \cite{06I}.  

\noindent 51) RPM's are a toy model for robustness issues: what if some degrees of freedom are ignored,  
along the lines in which Kucha\v{r} and Ryan \cite{KR} question whether Taub 
microsuperspace  sits stably inside Mixmaster as regards making QM predictions (itself a toy model of 
whether studying minisuperspace might be fatally flawed due to omitting all of the real universe's 
inhomogeneous modes).  
The RPM (or, for that matter, molecular) counterpart of this is rather easier to investigate. 

\noindent 52) Further arrow of time issues \cite{AOT} could conceivably be shared between RPM's and GR.

\mbox{ }

\noindent As a final non-analogy, I do not know of any meaningful Hartle--Hawking type condition on 
$\Psi$ for RPM's.

\section{Forms for the Laplacian}

The energy/Hamiltonian constraint's (\ref{EnCo}, \ref{HamCo})
$N^{\Gamma\Delta}(Q^{\Lambda})P_{\Gamma}P_{\Delta}$ term in general presents an operator ordering 
problem.  
One desirable condition pointed out by DeWitt \cite{DeWitt57} is for one's operator ordering to be 
invariant under coordinate changes. 
However, this does not fix ordering uniquely: the 1-parameter family $D^2 - \xi\, \mbox{Ric}(M)$ 
of $\xi$-orderings all obey this condition.  
This contains the Laplacian 
\beq
D^2 = \frac{1}{\sqrt{M}} \frac{\pa}{\pa {\cal Q}^{\Gamma}}
\left\{
\sqrt{M}N^{\Gamma\Delta}\frac{\pa}{\pa{\cal Q}^{\Delta}}
\right\}
\label{Lap}
\eeq
as a particular subcase, and also the conformally invariant operator 
$C^2 = D^2 - \xi_{\sc}\mbox{Ric}(M)$ with $\xi_{\sc} = \{k - 2\}/4\{k - 1\}$ as another particular 
subcase.  
I argue in favour of the latter from relational first principles in \cite{Banal} (previous 
arguments/uses of this ordering are e.g. in \cite{Magic, ConfOrder}).
In the present paper, I wish to consider another aspect of the operator ordering issue: that 
constraining before quantizing (the reduced or relationalspace approaches), constraining after 
quantizing (the Dirac approach) and considering the relational part of Newtonian Mechanics give 
different Laplacians (and so different $\xi$-operators also).

For N-stop metroland, $D^2_{\sN\sM} = D^2_{\sr\se\sd-\sr\se\sll} = 
D^2_{\sN\sM}|_{\sE-\sr\se\sll \, \sp\sa\sr\st}$, and conformal and $\xi$ operators throughout this 
Euclidean RPM example are the same since Ric($M$) = 0.  
Also, $C^2_{\sN\sM}|_{\sS-\sr\se\sll \, \sp\sa\sr\st} = D^2_{\sN\sM}|_{\sS-\sr\se\sll \, \sp\sa\sr\st} 
= D^2_{\sN\sM}|_{\pa/\pa\iota = 0} = \iota^2D^2_{\mathbb{S}^{\tN - 2}} = D^2_{\sr\se\sd-\sr\se\sll} 
\neq C^2_{\sr\se\sd-\sr\se\sll}$, though the last inequality is but by a constant which can 
be absorbed into a redefinition of the energy.  
[E- , S- and NM stand for Euclidean, similarity and Newtonian Mechanics.]

Also, 
\beq
D^2_{\sN\sM}|_{\sE-\sr\se\sll\, \sp\sa\sr\st} = D^2_{\sN\sM}|_{\pa/\pa\iota = 0} = 
4I\left\{\frac{1}{I^2}\frac{\pa}{\pa I}\left\{I^2\frac{\pa}{\pa I}\right\} + 
\frac{1}{I^2}\left\{\frac{1}{\mbox{sin}\,\Theta}\frac{\pa}{\pa\Theta}
\left\{\mbox{sin}\,\Theta\frac{\pa}{\pa\Theta}\right\} + \frac{1}{\mbox{sin}^2\Theta}
\frac{\pa^2}{\pa\Phi^2}\right\}\right\} \mbox{ } ,
\eeq 
while 
\beq
D^2_{\sE-\sr\se\sd-\sr\se\sll} = 
4I\left\{\frac{1}{I^{3/2}}\frac{\pa}{\pa I}\left\{I^{3/2}\frac{\pa}{\pa I}\right\} + 
\frac{1}{I^2}\left\{\frac{1}{\mbox{sin}\,\Theta}\frac{\pa}{\pa\Theta}
\left\{\mbox{sin}\,\Theta\frac{\pa}{\pa\Theta}\right\} + \frac{1}{\mbox{sin}^2\Theta}
\frac{\pa^2}{\pa\Phi^2}\right\}\right\} \mbox{ } ,  
\eeq
so that these differ by $ 2\pa_I = \ttD/I$.
Now also $C^2_{\sN\sM} = D^2_{\sN\sM}$ but $C^2_{\sE-\sr\se\sd-\sr\se\sll} \neq 
D^2_{\sE-\sr\se\sd-\sr\se\sll}$ and this is now not just a shift by a constant.

Finally, $D^2_{\sN\sM}|_{\sS-\sr\se\sll\,\sp\sa\sr\st} = D^2_{\sN\sM}|_{\pa/\pa A = 0, \pa/\pa\iota = 0} 
= D^2_{\sS-\sr\se\sd-\sr\se\sll}$ (the above difference is now removed due to containing the new 
constraint as a factor), though also $C^2_{\sN\sM}|_{\sS-\sr\se\sll\,\sp\sa\sr\st} = 
C^2_{\sN\sM}|_{\pa/\pa A = 0, \pa/\pa\iota = 0} \neq C^2_{\sS-\sr\se\sd-\sr\se\sll}$, albeit the 
difference is again but an absorbable constant.  
(Here, $A$ is an absolute angle).

Thus, this Sec's issue can be viewed as (mostly) stemming from the nontriviality of the rotations.  
It can also be viewed as some evidence in favour of Laplacian rather than conformal operator ordering 
(an idea developed further in \cite{RedDir}).

The discrepancy between the `relational portion' of Newtonian Mechanics's Laplacian and that of the 
relationalspace-reduced approaches is due to the general formula for the Laplacian (\ref{Lap}) 
containing $\sqrt{M}N^{\Gamma\Delta}$ which, in the former case, contains extra contributions from the 
absolute part even within the `relational portion'.  
By this means, if absolute space is assumed, it leaves an imprint on the `relational portion'.  
The effect of this is very loosely analogous to the issue of Newton's bucket \cite{bucket} in the 
absolute versus relative motion debate (with differences in shape of wavefunctions/probability densities in place of 
differences in the shape of the surface of the water).  
Also note that the imprint of a 3-$d$ absolute space is different from that of a 2-$d$ absolute space; 
the former is more complicated.    
It is in this way that the unusual operator ordering for the 3-body problem in the Molecular Physics 
literature (e.g. \cite{Iwai, PP87, Mead92} and the much earlier but non-geometrical \cite{Zick}) can be 
understood. 
E.g. from following through the chain of coordinate transformations in splitting 3-$d$ space into 
absolute and relative parts (rather than eliminating, or never starting with, the absolute part), the 
radial part Schr\"{o}dinger equation comes out as $\pa_I\mbox{}^2 + 5I^{-1}\pa_I$, while purely 
relational considerations give the sphere with reflexive identification, for which the far more usual 
$\pa_I\mbox{}^2 + 2I^{-1}\pa_I$ arises.  
[In 3-$d$ the shape part also differs between the split of Newtonian Mechanics and the relationalspace 
approach; see e.g. \cite{Iwai,PP87,Mead92,Zick} for various coordinatizations of the former's Laplacian.] 
%

\noindent 
Note I. The two situations are physically distinct (a molecule that is a small part of the universe 
versus a molecule that is a whole-universe model). 
Nevertheless, I would be interested to know whether the form used in Molecular Physics has detailed 
experimental confirmation in the former context.

\noindent
Note II. The Molecular Physics literature does not, as far as I have seen, make a connection between 
this ordering issue and the absolute versus relative motion; I mention here that the extent to which 
Aharonov and Kaufherr's \cite{Aharonov} work is related both to the situation in Molecular Physics {\sl 
and} to the above absolute versus relative motion debate issue represents interesting work in progress.  
 
\noindent 
Note III.  The purely relational operator ordering that I adopt has the added theoretical advantages of 
having 

\noindent (well-)known solutions in a number of cases, and of being in line with a number of authors' 
choice of operator ordering for the Wheeler--DeWitt equation.

\section{Do RPM's have configuration space monopole issues?}

Monopole issues are known to affect classical, and in particular, quantum-mechanical, study of a system.
E.g. consider a charged particle in 3-$d$ [for which Cartesian coordinates are $\underline{x} = (x^1, x^2, 
x^3)$ in Cartesian coordinates and ($r$, $\theta_{\sss\sp}$, $\phi_{\sss\sp}$) in spherical polar 
coordinates, 
with corresponding mechanical momentum $\underline{p}$, $m$ is its mass and $e$ is its charge] in the 
presence of a Dirac monopole \cite{DirMon, Dirac48, YW} of monopole strength $g$, corresponding to field 
strength 
\beq
F_{\beta\gamma} = \epsilon_{\alpha\beta\gamma}g{x^{\alpha}}/{r^3} \mbox{ } .  \label{FiSt}
\eeq
If one looks for a vector potential $\underline{A}$ corresponding to this in just the one chart, one 
finds that it is singular somewhere -- the monopole has a Dirac string emanating from it in some 
direction or other.  
However, there is no physical content in the direction in which it emanates, and one can avoid having 
such strings by using more than one chart (each chart's choice of Dirac string lying outside the chart). 
One such choice is to have an 
N-chart $\{\theta_{\sss\sp}, \phi_{\sss\sp} \mbox{ } | \mbox{ } 
0 \leq \theta_{\sss\sp} \leq \pi/2 + \epsilon\}$, and an 
S-chart $\{\theta_{\sss\sp}, \phi_{\sss\sp} \mbox{ } | \mbox{ } 
\pi/2 - \epsilon \leq \theta_{\sss\sp} \leq \pi\}$, with 
the vector potential in each of these being given by 
\beq
\underline{A}^{\sN} \cdot \d\underline{x} =  g\{x^1 \d x^2 - x^2 \d x^1\}/{r\{r + x^3\}} \mbox{ } , \mbox{ } \mbox{ } 
\underline{A}^{\sS} \cdot \d\underline{x} =  g\{x^1 \d x^2 - x^2 \d x^1\}/{r\{r - x^3\}} \mbox{ } .
\eeq 
Then the classical Hamiltonian for the charged particle is built from the canonical momentum combination 
$\underline{p} - e\underline{A}$ :
\beq
\fH = \{\underline{p} - e\underline{A}\}^2/2m + V(\underline{x}) \mbox{ } ,
\eeq
with $\underline{A}$ taking the above monopole form, i.e. $\underline{A}^{\sN}$ in the N-chart and 
                                                          $\underline{A}^{\sS}$ in the S-chart.  
Then the position and the canonical momentum are good Hermitian operators, but next in looking to form 
$SO$(3) objects, $\underline{A}$'s positional dependence complicates the commutation relations; this 
necessitates, beyond what is usual in angular momenta, the introduction of an extra term:  
\beq
\underline{L}^{\se\sx\st\se\sn\sd\se\sd} = 
\underline{x} \cr \{\underline{p} - e\underline{A}\} - q\underline{x}/r^3 \mbox{ } ,   
\mbox{ } \mbox{ for } \mbox{ } 
q = eg = n\hbar c/2 \mbox{ }  
\eeq
(the last equality being the Dirac quantization condition).  
Next, $L_{\sT\so\st\sa\sll}^{\se\sx\st\se\sn\sd\se\sd} = \sum_{\alpha = 1}^{3}
\{L_{\alpha}^{\se\sx\st\se\sn\sd\se\sd}\}^2 = 
\{\underline{x} \cr \{ \underline{p} - e\underline{A}\}\}^2 + q^2$, and 
$L_3^{\se\sx\st\se\sn\sd\se\sd}\Psi = \{-i\pa_{\phi_{\ts\tp}} - q\}\Psi$ in the N-chart and 
$L_3^{\se\sx\st\se\sn\sd\se\sd}\Psi = \{-i\pa_{\phi_{\ts\tp}} + q\}\Psi$ in the S-chart, so $\Psi$ 
has an exp$(i\{\nm \pm q\}\phi_{\sss\sp})$ factor rather than an exp$(i\nm\phi_{\sss\sp})$ factor, with 
integer m $\pm$ $q$. 
Consequently, the azimuthal part of the Schr\"{o}dinger equation then picks up two extra terms:
\beq
-  \{\mbox{sin}\,\theta_{\sss\sp}\}^{-1}
\{\mbox{sin}\,\theta_{\sss\sp}\Psi,_{\theta_{\sss\sp}}\},_{\theta_{\sss\sp}}
+  \{\mbox{sin}\,\theta_{\sss\sp}\}^{-2}\{\nm + q\,\mbox{cos}\,\theta_{\sss\sp}\}^2\Psi  
= \{\nl\{\nl + 1\} - q^2\}\Psi 
\mbox{ } ,
\eeq 
and so that `monopole harmonics' replace the ordinary spherical harmonics (\cite{YW}; e.g. \cite{ML97} 
has a 2-$d$ counterpart of closer relevance to the present paper).  
N.B. however how this working collapses in the case of an uncharged particle: the monopole is then not 
`felt' so one has the mathematically-usual particle Hamiltonian, the mathematically-usual form of the 
$SO$(3) operator, the mathematically-usual Schr\"{o}dinger equation and, for central potentials (i.e. 
potentials depending on the radial variable $r$ alone), the mathematically-usual spherical harmonics.  
This difference from \cite{08II} is one reason why \cite{08III} has had to wait for the below resolution.    

\mbox{ }

Now, monopole issues involving the triple collision are somewhat well-known to occur in 3-body problem 
configuration spaces.   
In the case of RPM's, is the above an indication of somewhat unusual mathematics arising analogously to in the above Dirac monopole 
considerations?

I begin by considering this for unoriented scaled triangleland in the Newtonian context of this sitting 
inside absolute space.
Firstly, the connection involved in this case is Guichardet's \cite{Guichardet} rotational connection 
(c.f. SSec 1 of Appendix).  
This is indeed a nontrivial connection as it has a nontrivial field strength (\ref{nontri}).  
In this case, passing to e.g. $\mbox{Dra}^{\Delta}$ coordinates gives precisely the Dirac monopole 
(\ref{FiSt}) under the correspondence $\mbox{Dra}^{\Delta}$ for $x^{\alpha}$, clustering (1)'s D in the 
role of N and its M in that of S.

Next, the Hamiltonian is (for relational space coordinates ${\cal Q}^{\Delta}$ with corresponding 
mechanical momenta ${\cal P}_{\Delta}$)  
\beq
\fH = \underline{L}\underline{\underline{I}}^{-1}\underline{L} + 
g^{\Gamma\Delta}\{{\cal P}_{\Gamma} - \underline{L}\cdot\underline{A}_{\Gamma}\}
                \{{\cal P}_{\Delta} - \underline{L}\cdot\underline{A}_{\Gamma}\}/2 
+ V({\cal Q}^{\Lambda}) \mbox{ } . 
\label{bigham}
\eeq
From here, monopole effects would spread into the Schr\"{o}dinger equation and its solutions.

However the key point is that the case of interest in RPM's is the one with {\sl zero total angular 
momentum} $\underline{L} = \uttL = 0$, for which, analogously to the special case of an uncharged 
particle in a Dirac monopole field, the Hamiltonian (\ref{bigham}) collapses to a much simpler form, 
\beq
\fH = g^{\Gamma\Delta}{\cal P}_{\Gamma}{\cal P}_{\Delta}/2 + V({\cal Q}^{\Lambda})
\eeq
corresponding to `not feeling' the monopole.  
Thus monopole effects do not enter the Schr\"{o}dinger equation and its solutions in this way (nor does 
a distinction between mechanical and canonical momentum change the form of the relative rational 
momentum operator in this case of zero total angular momentum).
Thus the associated $SO$(3) mathematics is standard, and, in the case of a `central' potential energy 
$V = V(I \mbox{ alone})$, the angular part of the Schr\"{o}dinger equation does indeed give the spherical 
harmonics (determining this requires additional working due to operator ordering issues, which is 
provided in the next Sec).   
Thus the scalefree RPM study of these in \cite{08II, +tri} turns out to be reuseable in the study of 
the scaled triangle as the shape part of the shape-scale split.

Note furthermore that if we choose, rather, oriented shapes (much more common in the literature due to 
the bias that the real world is 3-$d$ and one can therefore in this setting perform out-of the-plane 
rotations on a planar system), the monopole is not now of Dirac-type in 3-$d$ space but rather Iwai's 
monopole \cite{Iwai} in 3-$d$ half-space.   
The mathematics remains similar to that of the Dirac monopole in any given chart and gauge, however now 
other choices of string are more convenient \cite{Iwai, LR} (and the flux is halved due to involving 
half as much solid angle as before).  
In particular, the Hamiltonian remains of the form (\ref{bigham}) and the above simplification to this 
and to the subsequent QM for zero total angular momentum continues to apply (now involving some 
`half-sphere harmonics' rather than whole-sphere ones).

For more than three particles, the above simplifying effect of being in a $\uttL = 0$ theory (a useful 
result e.g. toward developing the quadrilateralland model), though I have not as yet considered in these 
cases whether the $g_{\Gamma\Delta}$ coincides with the metric obtained from first principles/from 
reduction of \cite{BB82}-type formulations.


Given that we straightforwardly eliminated the translations early on in this paper's treatment, I should 
next reassure the reader that, if these were left behind, they would not bring about a significant 
analogous connection to that brought about by the rotations.
This is because, as a further manifestation of the mathematical simpleness of the translations, their 
analogue of the Guichardet connection form has trivial field strength (SSec 2 of Appendix).  
Thus we can be sure that for scaled N-stop metroland (involving at most translations), one does not 
have monopole effects to worry about.    
This protects my treatment of N-stop metroland as `ordinary Euclidean space physics', among the 
coordinate systems for which the ((ultra)spherical) polar coordinates have a particularly lucid 
scale-shape interpretation.

Next, in the case of scalefree theories, can the dilations cause analogous monopole effects?   
I establish in SSec 2 of the Appendix that this is not the case because their analogue of the Guichardet 
connection form also has trivial field strength.   
Thus combining this result and the above rotational results, scalefree theories carry no vestiges of 
the excluded maximal collision through its action as a monopole (without establishing this, one could 
have feared e.g. that the shape spaces could contain a bad point of gauge-dependent position arising 
from the corresponding relationalspace's gauge and chart choice's Dirac string casting a `shadow' on the 
shape space at the point intersection in relationalspace between the Dirac string and the surrounding 
shape space, c.f. \cite{Dirac48}).
[Finally, translations, rotations and dilations do not interfere with each other if treated together 
in this sense.]

\section{What potentials to use in RPM's?}

SRPM has a lot of potential freedom (all potentials homogeneous of degree 0 are allowed), 
while ERPM can have any potential at all.
Moreover, what is effectively ERPM (the zero total angular momentum portion of Newtonian Mechanics) 
describes much of later-universe physics \cite{BGS03}.  
There is however a gap in that doing {\it Quantum} Cosmology isn't exactly later-universe physics, 
albeit Kucha\v{r} has also argued \cite{K99} that Halliwell and Hawking's perturbative treatment of 
inhomogeneous GR is likewise only a model of Quantum Cosmology, quantum fluctuations being far smaller 
than the classical universe's inhomogeneities.  
In models with inflation, this criticism may at least in part be circumvented.

Hitherto, the RPM program has been using HO potentials (or HO-like ones for SRPM, for which HO 
potentials themselves are disallowed by the homogeneity requirement).   
These are $\iota^2\{A + B\,\mbox{cos}\,2\varphi\}$ for 3-stop metroland, 

\noindent 
$\iota^2\{A + B\,\mbox{cos}\,2\theta + C\,\mbox{sin}^2\theta\,\mbox{cos}\,2\phi\}$ for 4-stop metroland,
and $I^2\{A + B\,\mbox{cos}\,2\Theta + C\,\mbox{sin}^2\Theta\,\mbox{cos}\,2\Phi \}$
for triangleland in the flat banal representation, which is also accompanied by a $E/4I$ term.  
These include well shapes about poles ($C$ = 0 case) \cite{08I, AF, +tri}.  
With scale added, c.f. \cite{08I} for triangleland: one has an infinite well or an infinite barrier 
depending on the sign of $E$.  
The scale or `radial direction for 3- and 4-stop metroland is a semi-finite well about the origin ($A > 
0$) or a semi-finite barrier ($A < 0$), though the latter does not have a multi-spring interpretation.  
Adding a ${\cal R}_{\sT\so\st\sa\sll}$ effective term adds spokes to the wells, including giving the 
familiar hydrogenic shape in combination with the downturning sign of $E/I$.

However, it is a main point of this paper to use the analogy between Mechanics and Cosmology to broaden 
the range of potentials under consideration and pinpoint ones which parallel Classical and Quantum 
Cosmology well.

\subsection{Ordinary 1-particle Mechanics -- Cosmology analogy}

Strong parallels between the Newtonian dynamics of a large dust cloud and of Einsteinian dust cosmology 
have been known for quite some time (see e.g. \cite{MMc}).  
For the present paper's purposes, enough of these parallels (\cite{HarMTWPee}) survive the introduction 
of a pressure term, as follows.   
%

Cosmology (in $c = 1$ units) has the Friedmann equation
\beq
H^2 \equiv \left\{\frac{\dot{a}}{a}\right\}^2 = - \frac{k}{a^2} + \frac{8\pi G\rho}{3} + \frac{\Lambda}{3} 
\mbox{ } . 
\eeq
Here, $a$ is the scalefactor of the universe, $H$ is the Hubble quantity, and the matter energy density 
$\rho \propto a^{-3}$ for dust or $\propto a^{-4}$ for ultrarelativistic `radiation fluid' from solving energy-momentum 
conservation (one gets the Friedmann, Raychaudhuri and energy-momentum conservation equations, but the last two 
are not independent, so one can solve either; conditions like `radiation' and `dust' then give an equation of 
state by which energy-momentum conservation can be solved thus).  
Also, $G$ is the gravitational constant, $k$ is the spatial curvature which is without loss of 
generality normalizable to 1, 0 or --1, and $M$ is the mass enclosed up to the radius $a(t)$. 

\noindent 
On the other hand, conservative mechanics of one particle in 3-$d$ and in spherical polar coordinates, 
has an energy constraint, which, divided by $r^2/2$ takes the form 
\beq
\left\{\frac{\dot{r}}{r}\right\}^2 = \frac{2E}{r^2} - \frac{(\mbox{shape term})}{r^2} 
                                                    - \frac{2V(r, \mbox{shape})}{r^2} \mbox{ } .
\eeq
Thus there is an analogy between Friedmann {\sl after solution of Raychaudhuri or energy-momentum conservation} 
and mechanics.  
In this,
 -- $2E$ plays the same role as curvature (and can again without loss of generality be set to 
1, 0 or --1).
The shapefree part of $1/r$ (`Newtonian') potentials are likewise analogous to dust. 
The shapefree part of $r^2$ (`HO') potentials are analogous to the cosmological constant.

The `central' or `axisymmetric' term has the form of negative density (`wrong sign') radiation fluid.
In the cosmological context, this still has $P = \rho/3$ equation of state (for $P$ the pressure),  
but $\rho$ is negative.  
Consequently it violates all energy conditions, making it likely to be unphysical. 
This is not particularly desirable cosmologically, albeit I) this can lead to singularity theorem 
evasion -- `bounce' models.   
II) Nor has this undesirability stopped such matter appearing as `dark radiation' \cite{DarkRad} in models 
that rest on more exotic, geometrically complicated scenarios such as brane cosmology \cite{DarkRad, 
AT05} being able to possess what appears to be energy condition violation from the 4-$d$ spacetime 
perspective due to projections of higher-dimensional objects.
Moreover, this `wrong-sign radiation' term is entirely expected in mechanics (it is the centrifugal 
barrier preventing collapse so that one indeed expects a `bounce' rather than collapse to singular 
behaviour like a Friedmann model's Big Bang singularity). 
This difference in sign can be seen as originating from an important limitation in the 
Mechanics--Cosmology analogy: expansion in GR contributes negatively to kinetic energy while it does 
so positively in mechanics.
In this paper, this can be absorbed by reversing all the (pseudo)potential signs and the sign of the 
overall energy (-- 2$E$ in place of $k$); note that the shape terms in the RPM kinetic energy are of 
the same sign as the scale term, while in GR they are of opposite sign; it is from this that 
the radiation-like term gets the `wrong sign'. 
Finally, note that one can suppress this term if one's model's relative rational momentum is 
zero/small/swamped by the following contribution to the potential.

Ordinary radiation fluid, whether because one wishes for models paralleling universes for which a such 
is significant, or so as to cancel/outweigh the wrong-sign radiation term that is {\sl mechanically} 
well-motivated, can be brought about by inclusion of a $1/r^2$ (`conformally invariant') potential term.
This is the conformally invariant potential term $1/r^2$, which is quite well-studied in Classical and 
Quantum Mechanics.

What we want is the parallel of this analogy for reduced 1- and 2-$d$ RPM's, as these have further 
closed-universe and GR-like connotations.   
Scale dynamics is to dominate (heavy slow motion) so we use RPMs' freedom of form of potential to match 
scale dynamics to forms reasonably in line with cosmological scale dynamics.     
There is then a regime in which shape dependence is small and then big shapefree terms appear by 
expanding the potential contributions.

\subsection{RPM -- Cosmology analogy}

\begin{tabbing}

\hspace{0.1in}\underline{\footnotesize{scalefactor $a$}} \hspace{1.4in} \= \footnotesize{\underline{hyperradius scale $\iota$}}  
\hspace{1.1in} \= \footnotesize{\underline{moment of inertia scale $I$}} \\

\footnotesize{22) Friedmann equation} \hspace{0.05in} \>         
$\left\{\frac{\dot{\iota}}{\iota}\right\}^2 =  
                                             \frac{2E}{\iota^2} 
                                             - \frac{{\cal D}_{\tT\to\ttt\ta\tl}}{\iota^4} 
                                             - \frac{2V(\iota, \, \sss\sh\sa\sp\se)}{\iota^2}$        \> 

$\left\{\frac{\dot{I}}{I}\right\}^2 =  \frac{2E}{I^3} - \frac{{\cal R}_{\tT\to\ttt\ta\tl}}{I^4} 
                                                     - \frac{2V(I, \, \sss\sh\sa\sp\se)}{I^3}$        \\

\footnotesize{23)  spatial curvature term $k$}               \>  \footnotesize{$-2E$}  \>  \footnotesize{$2A$}                    \\
\footnotesize{24)  cosmological constant term $\Lambda/3$}   \>  \footnotesize{$-2A$ from Hooke-type potential} \>   
                                    \footnotesize{surviving term from $r_{IJ}^6$ potentials, with coefficient $2S$} \\
\footnotesize{25)  dust term coefficient $2GM$}                   \>  \footnotesize{$-2K$ from Newtonian potential}  \>  \footnotesize{$2E$} \\
\footnotesize{26)  radiation term  coefficient $2GM/a^4$}          \>  \footnotesize{$2R$ from conformally-invariant potential} \> 
\footnotesize{$2R$ from conformally-invariant potential} \\

\footnotesize{a)\mbox{ } Wrong-sign radiation term}   \>   \footnotesize{$-{\cal R}_{\tT\to\ttt\ta\tl}$} \> 
\footnotesize{$-{\cal R}_{\tT\to\ttt\ta\tl}$} \\
\footnotesize{Overall right-or-wrong sign radiation term} \> 
\footnotesize{$2R - {\cal R}_{\tT\to\ttt\ta\tl}$} \>  \footnotesize{$2R - {\cal R}_{\tT\to\ttt\ta\tl}$} \\

\end{tabbing}

\noindent
Note I) a) is another analogy, now with with ordinary mechanics, as expalained above. 

\noindent 
Note II) The second column applies to 1-$d$, or 2-$d$'s $\mathbb{CP}^k$ presentation, while for 2-$d$'s 
$\mathbb{S}^2$, one gets the third column's analogy.  
The spherical triangeland analogy is of limited use due to not extending to higher N-a-gonlands; one use 
for it is use that it allows the shape part to be studied in $\mathbb{S}^2$ terms which more closely 
parallel the Halliwell--Hawking \cite{HallHaw} analysis of GR inhomogeneities over $\mathbb{S}^3$.  

\noindent 
Note III) In 22), note that the energy equation is analogous to the Friedmann equation {\sl after} use 
of the energy-momentum conservation equation or the Raychaudhuri equation.  

\noindent
Note IV) To match the $k$ = --1, 0 or 1 convention, I use the corresponding scaling freedom in RPM's to 
set $-2E$ =  --1, 0 or 1 in the standard analogy, and $2A$ = --1, 0 or 1 in the $\mathbb{S}^2$ 
presentation of triangleland's analogy (which freedom remained unused in \cite{08I, 08II, +tri}); this 
amounts to the other coefficients being redefined by that constant factor; of course, this does not 
change any results).

\noindent Note V) 
In the spherical triangleland analogy, the Newtonian $1/|r_{IJ}|$ type potentials that one might 
consider to be mechanically desirable to include produce $1/I^{7/2}$ terms, which are analogous to an 
effective fluid with equation of state $P = \rho/6$ (for $P$ the pressure) i.e. an interpolation 
`halfway between' radiation fluid and dust. 
This is physically reasonable for a cosmology: it does not violate any energy conditions and is sensible 
as a rough model of a mixture of dust and radiation as is believed to have been present when the 
universe was around 60000 years old.

\noindent Note VI) 
Throughout, these analogies are subject to any shape factors present being slowly varying/expandible with a dominant constant leading 
term, at least in some region of interest; because of this, the analogy is between exact isotropic 
cosmology solutions and approximate solutions for the mechanics of scale. 
Moreover, N.B. that isotropic cosmology itself similarly suppresses small anisotropies and 
inhomogeneities, so that exact solutions thereof are really approximate solutions for more realistic 
universes too.

\subsection{Many-particle Mechanics -- Cosmology analogy}

In considering a large number of particles, another way in which shape could be at least approximately 
negligible arises \cite{BGS03}: through its overall averaging out to produce a highly radial problem (in 
a factorization into a cosmology-like scale problem and a shape problem).  
In the case of dust in 3-$d$, the many Newtonian gravity potential terms average out to produce the 
effective dust, and one's equations are split into the standard dust cosmological scale equation (good 
for most cosmological aspects of the later universe -- microlensing requires exception or additional 
prescriptions in this setting) and the central configuration problem for shape, which is also 
well-known.\footnote{Albeit this is well-known 
under somewhat different circumstances: the Celestial Mechanics literature considers the few-particle 
case (see e.g. \cite{Moeckel} for an introduction and further references), while the 
late universe cosmological interest is in the large-particle limit \cite{BGS03}.}
%
The Newton--Hooke problem, amounting to cosmology with dust and cosmological constant, has also been 
studied in a somewhat similar context (see e.g. \cite{GP}).  
It is an interesting question to me whether the averaging out to produce a radial equation and a shape 
equation occurs for other power-law potentials and their superpositions and whether any of the resulting 
shape problems are of mathematical form that has been substantially studied before.

\subsection{Choice of research direction}

I currently give preference to analogue models of realistic cosmologies over models that are 
well-known from the Mechanics/Molecular Physics literatures, as the virtues of RPM's that I intend to 
exploit are in their being toy models for whole universe GR, and I argue that using up RPM's freedom 
of potential to make their scale dynamics approximately match that of realistic GR cosmology models is a 
useful way to further enhance the analogy. 
Now, instead of such a scale dynamics being coupled to a complicated GR structure formation process, it 
is coupled to a simpler structure formation process that is finite and tied to well-studied 
configuration space geometry (e.g. $\mathbb{S}^{\sN - 2}$ or $\mathbb{CP}^{\sN - 2}$), which, 
nevertheless is still of value as regards the investigation of a number of conceptual issues such as the 
Problem of Time in Quantum Gravity.   
The above analogies then include a reasonable justification of various of my previous papers using HO's 
\cite{TriCl,08I,08II,MGM,AF,+tri} as well as suggesting further potentials of interest.  
This is brought out somewhat more clearly still by considering what the solutions to various Friedmann 
equations are, and what then are the corresponding solutions to N-stop metroland and triangleland RPM 
analogue Friedmann equations.

Subsequent papers \cite{scaleQM, 08III, RedDir} will cover analogies at the level of the quantum 
Friedmann equations, and use above and below classical work as the first-approximation H-part of the 
semiclassical approach to the Problem of Time.  
These papers will apply further selection filters as regards which potentials are 
theoretically-desirable.

\section{RPM-Cosmology analogy: simple classical solutions}

\subsection{A range of standard GR isotropic cosmology solutions}

\noindent ${\cal A}$) Models with spatial curvature and cosmological constant are as follows (see e.g. 
\cite{Rindler}).     

\noindent
i)   $k = 0$, $\Lambda = 0$ is $a$ constant: a static universe.  

\noindent
ii)  $k = -1$, $\Lambda = 0$ is $a = t$.

\noindent
iii) $k = 1$ and $\Lambda \leq 0$ is impossible.  

\noindent
iv)  $k = 0$ and $\Lambda > 0$ is $a = \mbox{exp}(\sqrt{\Lambda/3}t)$.

\noindent
v)   $k = -1$ and $\Lambda > 0$ is $a = \sqrt{3/\Lambda}\,\mbox{sinh}(\sqrt{\Lambda/3}t)$.  

\noindent
vi)  $k = 1$ and $\Lambda > 0$ is $a = \sqrt{3/\Lambda}\,\mbox{cosh}(\sqrt{\Lambda/3}t)$. 

\noindent
Note that iv) to vi) are all de Sitter/inflationary type models. 

\noindent
vii) $k = -1$ and $\Lambda < 0$ is $a = \sqrt{-3/\Lambda}\,\mbox{sin}(\sqrt{-\Lambda/3}t)$ 
-- a `Milne in anti de Sitter' oscillating solution.  


\noindent ${\cal B}$)
Models with spatial curvature and dust are as follows.    

\noindent
i)   $k = 1$ is the well-known cycloid solution.  

\noindent
ii)  $k = 0$ $a = \{9GM/2\}^{1/3}t^{2/3}$. 

\noindent
iii) $k = - 1$ is the also well-known hyperbolic analogue of the cycloid. 


\noindent ${\cal C}$) Models with radiation and spatial curvature include the following solutions 
\cite{Wald}.    

\noindent 
i) $k = 1$: $a = \sqrt{2GM}\{1 - \{1 - t/\sqrt{2GM}\}^2\}^{1/2}$ (the Tolman model).  

\noindent 
ii) $k = 0$: $a = \{8GM\}^{1/4}t^{1/2}$.

\noindent 
iii) $k = - 1$: $a = \sqrt{2GM}\{\{1 + t/\sqrt{2GM}\}^2 - 1\}^{1/2}$.  


Also note that 

\noindent
${\cal D}$) the case of $P = \rho/6$ is not exactly integrable except for $k = 0$, in which case the solution is 
$a = \{2GM\}^{13/21}\{4t/3\}^{4/7}$.

\noindent Finally, I consider further combinations of the well-motivated potential terms. 
E.g., 

\noindent ${\cal E}$) The cosmologically standard model comprising dust, spatial curvature and cosmological 
constant is covered e.g. in \cite{Rindler, Edwards72}.    
Solutions of this include the Lema\^{\i}tre model, a  model in which the Big Bang tends to the Einstein 
static model, the Eddington--Lema\^{\i}tre model, and various oscillatory models including a bounce.  
Solutions for $k$ and $\Lambda$ both nonzero involve in general elliptic integrals.  
Some subcases that are solvable in basic functions are 

\noindent
i)   $k = 0$, $\Lambda > 0$, solved by   
$a = \{\{3GM/\Lambda\}\{\mbox{cosh}(\sqrt{3\Lambda}t) - 1\}\}^{1/3}$, and 

\noindent
ii)  $k = 0$, $\Lambda < 0$, solved by   
$a = \{\{-3GM/\Lambda\}\{1 - \mbox{cos}(\sqrt{-3\Lambda}t)\}\}^{1/3}$.


\subsection{Further support from ordinary Mechanics}

`Wrong sign radiation' in Cosmology clearly corresponds via the Cosmology--Mechanics analogy to just the 
kind of effective potential term that one has for a centrifugal barrier, which is often present and 
well-studied in ordinary mechanics. 
Thus this case, while cosmologically unusual, does not lead to any unusual calculations either.  
In any case, such sign changes usually do not change exact integrability, but can change the qualitative 
behaviour in at least some regimes (consider e.g. trigonometric functions becoming hyperbolic functions 
under a sign change in some elementary integrals).  

\noindent With $E$ and overall angular momentum term (= wrong-sign radiation), ${\cal C}$)iv),v) are 
impossible cases, while ${\cal C}$vi) gives $r = \sqrt{t^2 + L^2}$.  

\noindent Some cases of this remain readily tractable if a Kepler--Coulomb potential term is added 
to this.  

\noindent The next 2 SSecs cover te different translation schemes for each of I) N-stop metroland and 
the $\mathbb{CP}^{\sN - 2}$ presentation of N-a-gonland and II) the $\mathbb{S}^2$ presentation of 
triangleland.

\subsection{N-stop metroland analogue cosmology approximate scale equation solutions}

[These also apply to the $\mathbb{CP}^{\sN - 1}$ presentation of N-a-gonland, under 
${\cal D}_{\sT\so\st\sa\sll} \longrightarrow {\cal R}_{\sT\so\st\sa\sll}$]

\noindent ${\cal A}$) Models with energy and (upside-down) HO potentials' scale contribution are as follows.     

\noindent
i)   $E = 0$, $A = 0$ is $\iota$ constant: a static model universe.  

\noindent
ii)  $E = 1/2$, $A = 0$ is $\iota = t$.

\noindent
iii) $E = -1/2$ and $A \geq 0$ is impossible.  

\noindent
iv)  $E = 0$ and $A < 0$ is $\iota = \mbox{exp}(\sqrt{-2A}t)$.

\noindent
v)   $E = 1/2$ and $A < 0$ is $\iota = \{1/\sqrt{-2A}\}\mbox{sinh}(\sqrt{-2A}t)$.  

\noindent
vi)  $E = -1/2$ and $A < 0$ is $\iota = \{1/\sqrt{-2A}\}\mbox{cosh}(\sqrt{-2A}t)$. 

\noindent
vii) $E = 1/2$ and $A > 0$ is $\iota = \{1/\sqrt{2A}\}\mbox{sin}(\sqrt{2A}t)$.


\noindent ${\cal B}$) Models with energy and Newtonian potentials' scale contribution are as follows.   

\noindent
i)   $E = -1/2$ is a cycloid solution.  

\noindent
ii)  $E = 0$ is $\iota = \{-9K/2\}^{1/3}t^{2/3}$.  

\noindent
iii) $E = 1/2$ is a hyperbolic analogue of the cycloid. 


\noindent ${\cal C}$) Models with energy and scale-invariant potential terms have the following approximate 
heavy-scale solutions (the $R$ is but the constant lead term of an expansion in the shape variables).  
For $2R - {\cal D}_{\sT\so\st\sa\sll} > 0$ (corresponding to right-sign radiation in Cosmology), 

\noindent 
i) $E = -1/2$: $\iota = \sqrt{2R - {\cal D}_{\sT\so\st\sa\sll}}\{1 - \{1 - 
t/\sqrt{2R - {\cal D}_{\sT\so\st\sa\sll}}\}^2\}^{1/2}$.

\noindent 
ii) $E = 0$: $\iota = \{4\{2R - {\cal D}_{\sT\so\st\sa\sll}\}\}^{1/4}t^{1/2}$.

\noindent 
iii) $E = 1/2$: $\iota = \sqrt{2R - {\cal D}_{\sT\so\st\sa\sll}}
\{\{1 + t/\sqrt{2R - {\cal D}_{\sT\so\st\sa\sll}}\}^2 - 1\}^{1/2}$

\noindent For $2R - {\cal D}_{\sT\so\st\sa\sll} < 0$, including ${\cal D}_{\sT\so\st\sa\sll} \neq 0$, 
$R = 0$, and corresponding to wrong-sign radiation in Cosmology, 

\noindent 
iv) $E = -1/2$ is impossible, 

\noindent
v)  $E = 0$ is also impossible, and 

\noindent
vi) $E = 1/2$ gives $\iota = \sqrt{t^2 + {\cal D}_{\sT\so\st\sa\sll} - 2R}$.  


\noindent ${\cal E}$) Some further approximate (the $K$ is but the constant lead term of an expansion in the 
shape variables) heavy-scale solutions are as follows.  

\noindent i)   $E = 0$, with upside-down HO $A < 0$ and Newtonian potential terms, solved by    
$\iota = \{\{K/2A\}\{\mbox{cosh}(3\sqrt{-2A}t) - 1\}\}^{1/3}$.  

\noindent
ii)  $E = 0$, with HO  $A > 0$ and Newtonian potential terms, solved by $\iota = 
\{\{-K/2A\}\{1 - \mbox{cos}(3\sqrt{2A}t)\}\}^{1/3}$.

\subsection{Triangleland analogue cosmology approximate scale equation solutions}

\noindent ${\cal A}$) Models with (upside down) HO and $|r^{IJ}|^6$ potential terms are as follows (these are just 
approximate heavy solutions in cases with $S \neq 0$).

\noindent
i)   $A = 0$, $S = 0$ is $I$ constant: static universe.  

\noindent
ii)  $A = -1/2$, $S = 0$ is $I = t$.

\noindent
iii) $A = 1/2$ and $S \leq 0$ is impossible.  

\noindent
iv)  $A = 0$ and $S > 0$ is $I = \mbox{exp}(\sqrt{2S}t)$.

\noindent
v)   $A = -1/2$ and $S > 0$ is $I = \{1/\sqrt{2S}\}\mbox{sinh}(\sqrt{2S}t)$.  

\noindent
vi)  $A = 1/2$ and $S > 0$ is $I = \{1/\sqrt{2S}\}\mbox{cosh}(\sqrt{2S}t)$. 

\noindent
vii) $A = -1/2$ and $S < 0$ is $I = \{1/\sqrt{-2S}\}\mbox{sin}(\sqrt{-2S}t)$.  


\noindent ${\cal B}$)
Models with (upside down) HO and energy have the following solutions.  

\noindent
i)   $A = 1/2$ is a cycloid.  

\noindent
ii)  $A = 0$ $I = \{9E/2\}^{1/3}t^{2/3}$.  

\noindent
iii) $A = - 1/2$ is the hyperbolic analogue of the cycloid. 


\noindent ${\cal C}$) Models with conformally invariant potential and (upside down) HO include the following 
solutions. 
For $2R - {\cal R}_{\sT\so\st\sa\sll} > 0$ (corresponding to right-sign radiation in 
Cosmology), 

\noindent 
i) $A = 1/2$: $I = \sqrt{2R - {\cal R}_{\sT\so\st\sa\sll}}
\{1 - \{1 - t/\sqrt{2R - {\cal R}_{\sT\so\st\sa\sll}}\}^2\}^{1/2}$.

\noindent 
ii) $A = 0$: $I = \{4\{2R - {\cal R}_{\sT\so\st\sa\sll}\}   \}^{1/4}t^{1/2}$.

\noindent 
iii) $A = - 1/2$: $I = \sqrt{2R - {\cal R}_{\sT\so\st\sa\sll}}
\{\{1 + t/\sqrt{2R - {\cal R}_{\sT\so\st\sa\sll}}\}^2 - 1\}^{1/2}$.  

\noindent For $2R - {\cal R}_{\sT\so\st\sa\sll} < 0$, including ${\cal R}_{\sT\so\st\sa\sll} 
\neq 0$, $R = 0$, and corresponding to wrong-sign radiation in Cosmology, 

\noindent 
iv) $A = 1/2$ is impossible, 

\noindent
v)  $A = 0$ is also impossible, and 

\noindent
vi) $A = -1/2$ gives $I = \sqrt{t^2 + {\cal R}_{\sT\so\st\sa\sll} - 2R}$.  


\noindent ${\cal D}$) The model with Newtonian potentials and $E = 0$ has the approximate heavy solution 
$I = \{2E\}^{13/21}\{4t/3\}^{4/7}$ which parallels the flat cosmology with $P = \rho/6$.    


\noindent ${\cal E}$) Some further approximate heavy-scale solutions are as follows.  

\noindent
i) $A = 0$, $S > 0$ and E-term of suitable sign: 
$I = \{\{E/2S\}\{\mbox{cosh}(3\sqrt{2S}t) - 1\}\}^{1/3}$. 

\noindent
ii) $A = 0$, $S < 0$ and E-term of suitable sign: 
$I = \{\{-E/2S\}\{1 - \mbox{cos}(3\sqrt{-2S}t)\}\}^{1/3}$.

\subsection{Comments on these RPM solutions}

Note in particular that the cyclic trial models with HO mathematics ${\cal A}$vii) of \cite{AF, 08I, 
08II, +tri} do correspond to a known cosmology (Milne in anti de Sitter) and that having some 
upside-down HO's, rather (also readily tractable), is de Sitter/inflationary in character 
[${\cal A}$iv), v) and vi)].  
Other models parallel the dynamics of fairly realistic simple models of the early universe involving 
radiation, spatial curvature and cosmological constant type terms.

Models with further combinations of right-sign radiation conformally invariant potential term, energy, 
Newton--Coulomb term and 
(upside-down) HO term for N-stop metroland, and of conformally invariant potential terms, energy, 
(upside down) HO terms and $|r^{IJ}|^6$ potential terms for triangleland, can readily be obtained by 
applying the analogy to the following cosmological models. 
Models with radiation and spatial curvature and cosmological constant include a 
subcase of what is covered by Harrison \cite{Harrison67} and Vajk \cite{Vajk}.  
Models with all of radiation, dust, spatial curvature and cosmological constant are also a subcase of 
what is treated in Harrison \cite{Harrison67}, and also more explicitly by Coqueraux and Grossmann 
\cite{CG82} and by Dabrowski and Stelmach \cite{DS86}.
While, the analogy with ordinary mechanics covers combining a `wrong sign radiation' `central term'  
with these other terms (e.g. in the Newton--Hooke problem).

\section{Making $t$ the subject and semiclassical approach applications}

This is useful in the semiclassical approach to the Problem of Time as the provision of an emergent 
approximate timestandard therein.  
I discuss this for the N-stop metroland/$\mathbb{CP}^{\sN - 2}$ presentation of N-a-gonland (use $I$ in 
place of $\iota$ for the corresponding $\mathbb{S}^2$ presentation of triangleland).  
The Born--Oppenheimer and WKB ans\"{a}tze are $\Psi = \psi(\iota)|\eta(\iota, \mbox{shape})\rangle$, 
$\psi = \mbox{exp}(iW(\iota)/\hbar)$.  
Then the H-equation is $\langle\eta|\widehat{H}\{|\eta\rangle\psi\} = 0$ and the L-equation is 
$\{1 - |\eta\rangle\langle\eta|\}\hat{H}\{|\eta\rangle\psi\} = 0$.  
One then usually uses a Hamilton--Jacobi equation approximation to the H-equation.  
Moreover, this corresponds to an energy equation and hence to an analogue Friedmann equation [via 
$\pa \fW/\pa\iota$ (for $\fW$ the principal function) to $p_\iota$ to $\d \iota/\d t^{\se\sm}$], so that one can feed in a range of 
cosmologically-plausible scale H-dynamics, to each of which we can couple a simple-to-study L-dynamics 
of pure shape.  
I take the following possible and nontrivial cases to suffice for the moment.  
The $t^{\se\sm}$ here is the mechanically-natural one, which is approximately the same as the WKB one 
of the emergent semiclassical approach too (there is an infinite family of banally-related emergent 
times, of which the $t^{\se\sm}$ here is but one particular member.)

\begin{tabbing}

\underline{Model} \hspace{0.1in} \= \underline{N-stop metroland case} \hspace{2in} \=
\underline{triangleland case} \\

\noindent ${\cal A}$)ii) \>  $t^{\se\sm} = \iota$     \>  $t^{\se\sm} = I$ \\

\noindent ${\cal A}$iv) \>  $t^{\se\sm} = \{1/\sqrt{-2A}\} \mbox{ln}\,\iota   $ \>
$t^{\se\sm} = \{1/\sqrt{2S}\} \mbox{ln}\,I$ \\

\noindent  ${\cal A}$v) \>  $t^{\se\sm} = \{1/\sqrt{-2A}\} \mbox{arsinh}(\sqrt{-2A}\iota)  $ \>
$t^{\se\sm} = \{1/\sqrt{2S}\} \mbox{arsinh}(\sqrt{2S}I)  $ \\

\noindent  ${\cal A}$vi) \>  $t^{\se\sm} = \{1/\sqrt{-2A}\} \mbox{arcosh}(\sqrt{-2A}\iota)  $ \>
$t^{\se\sm} = \{1/\sqrt{2S}\} \mbox{arcosh}(\sqrt{2S}I)  $ \\

\noindent  ${\cal A}$vii)\> $t^{\se\sm} = \{1/\sqrt{2A}\}  \mbox{arcsin}(\sqrt{2A}\iota)$ \>
$t^{\se\sm} = \{1/\sqrt{-2S}\}  \mbox{arcsin}(\sqrt{-2S}I)$ \\

\noindent ${\cal B}$)ii) \> $t^{\se\sm} = \sqrt{-2/9K}\iota^{3/2}  $ \>
$t^{\se\sm} = \sqrt{2/9E}I^{3/2}  $ \\

\noindent ${\cal C}$)i) \> $t^{\se\sm} =  \sqrt{2R - {\cal D}_{\sT\so\st\sa\sll}} 
\{1 - \sqrt{1 - \iota^2/\{2R - {\cal D}_{\sT\so\st\sa\sll}}\}\} $ \>

$t^{\se\sm} =  \sqrt{2R - {\cal R}_{\sT\so\st\sa\sll}}
\{1 - \sqrt{1 - I^2/\{2R - {\cal R}_{\sT\so\st\sa\sll}}\}\} $ \\

\noindent  ${\cal C}$ii) \>  $t^{\se\sm} =  \iota^2/2\sqrt{2R - {\cal D}_{\sT\so\st\sa\sll}} $ \>

$t^{\se\sm} =  I^2/2\sqrt{2R - {\cal R}_{\sT\so\st\sa\sll}} $ \\

\noindent  ${\cal C}$iii) \> $t^{\se\sm} =  \sqrt{2R - {\cal D}_{\sT\so\st\sa\sll}}
\{\sqrt{1 + \iota^2/\{2R - {\cal D}_{\sT\so\st\sa\sll}}\} - 1\} $ \>

$t^{\se\sm} =  \sqrt{2R - {\cal R}_{\sT\so\st\sa\sll}}
\{\sqrt{1 + I^2/\{2R - {\cal R}_{\sT\so\st\sa\sll}}\} - 1\} $ \\

\noindent  ${\cal C}$vi) \>$t^{\se\sm} = \sqrt{\iota^2 - D_{\sT\so\st\sa\sll} + 2R}$ \>
$t = \sqrt{I^2 - R_{\sT\so\st\sa\sll} + 2R}$  \\

\noindent ${\cal D}$) \> \> $t^{\se\sm} = 3I^{7/4}/4\{2E\}^{13/12}$          \\

\noindent ${\cal E}$)i) \> $t^{\se\sm} =  \{1/3\sqrt{-2A}\}\mbox{arccosh}(1 + 2A\iota^3/K)  $ \>
$t^{\se\sm} =  \{1/3\sqrt{2S}\}\mbox{arccosh}(1 + 2SI^3/E)  $ \\

\noindent ${\cal E}$ ii) \> $t^{\se\sm} = \{1/3\sqrt{2A}\}\mbox{arccos}(1 + 2A\iota^3/K)  $ \>
$t^{\se\sm} = \{1/3\sqrt{-2S}\}\mbox{arccos}(1 + 2SI^3/E)  $ \\

\end{tabbing}

\noindent I note that all of the above approximate heavy-scale timefunctions are monotonic apart from 
the ${\cal A}$)vii) and ${\cal E}$)i) models', which nevertheless have a reasonably long era of 
monotonicity as regards modelling early-universe Quantum Cosmology.  
${\cal A}$)vii) and ${\cal E}$)i) have periods proportional to $1/\sqrt{A}$ for N-stop metroland, and to 
$1/\sqrt{S}$ for triangleland, each of which plays a role proportional to that of $1/\sqrt{\Lambda}$ in 
GR Cosmology.  
Also, ${\cal A}$)iv) hits zero scale at other than t = 0 (one might reset origin of time to deal with 
this).
Finally, in each case, ${\cal A}$)v) and ${\cal C}$vi) have a nonzero minimum size (c.f. Sec 8.5's 
discussion for the latter).

In the L-equation, we get 
\beq 
\frac{\hbar^2}{2}2N^{\iota\iota}\frac{\pa\psi}{\pa\iota}\frac{\pa|\eta\rangle}{\pa\iota} = 
\hbar^2\frac{i}{\hbar}N^{\iota\iota}\frac{\pa\psi}{\pa\iota}\frac{\pa|\eta\rangle}{\pa\iota} 
\mbox{ } \mbox{ which contains } \mbox{ } 
i\hbar p_{\iota}\frac{\pa|\eta\rangle}{\pa\iota} = 
i\hbar\frac{\pa\iota}{\pa t^{\se\sm}}\frac{\pa|\eta\rangle}{\pa\iota}
= i\hbar\frac{\pa|\eta\rangle}{\pa t^{\se\sm} } \mbox{ } . 
\eeq
This arises alongside $\{\hbar^2/\iota^2\}D_{\sss\sh\sa\sp\se}^2|\eta\rangle$.  
Thus a new timefunction $t^{\sr\se\sc}$ (rec for rectified) such that $\pa/\pa t^{\sr\se\sc} = 
\iota^{2}\pa/\pa t^{\se\sm}$ simplifies things, 
so I also tabulate and discuss that.
The constant reflects a freedom in choice of origin of $t^{\sr\se\sc}$.  


\begin{tabbing}

\underline{Model} \hspace{0.1in} \= \underline{N-stop metroland case} \hspace{2in} \=
\underline{triangleland case} \\

\noindent ${\cal A}$)ii) \>  
$t^{\sr\se\sc} = \mbox{const} - 1/t^{\se\sm}$ \> 
$t^{\sr\se\sc} = \mbox{const} - 1/t^{\se\sm}$ \\

\noindent  ${\cal A}$iv) \>  
$t^{\sr\se\sc} = \mbox{const} - \mbox{exp}(-2\sqrt{-2A}t^{\se\sm})/2\sqrt{-2A}$ \>
$t^{\sr\se\sc} = \mbox{const} - \mbox{exp}(-2\sqrt{2S}t^{\se\sm})/2\sqrt{2S}$ \\

\noindent  ${\cal A}$v)  \>  
$t^{\sr\se\sc} = \mbox{const}  - \mbox{coth}(\sqrt{-2A}t^{\se\sm})/\{-2A\}^{3/2}$ \>
$t^{\sr\se\sc} = \mbox{const}  - \mbox{coth}(\sqrt{2S}t^{\se\sm})/\{2S\}^{3/2}$ \\

\noindent  ${\cal A}$vi) \>  
$t^{\sr\se\sc} = \mbox{const} + \mbox{tanh}(\sqrt{-2A}t^{\se\sm})/\{-2A\}^{3/2}$ \>
$t^{\sr\se\sc} = \mbox{const} + \mbox{tanh}(\sqrt{2S}t^{\se\sm})/\{2S\}^{3/2}$ \\

\noindent  ${\cal A}$vii)\> 
$t^{\sr\se\sc} = \mbox{const} - \mbox{cot}(  \sqrt{2A}t^{\se\sm}  )/\{  \sqrt{2A}\}^{3/2}$   \>
$t^{\sr\se\sc} = \mbox{const} - \mbox{cot}(  \sqrt{-2S}t^{\se\sm}  )/\{  \sqrt{-2S}\}^{3/2}$   \\

\noindent ${\cal B}$ii) \> 
$t^{\sr\se\sc} = \mbox{const} - \{\frac{-2}{K}\}^{2/3}\frac{1}{{3t^{\te\tm}}^{1/3}}$ \>
$t^{\sr\se\sc} = \mbox{const} - \{\frac{2}{E}\}^{2/3}\frac{1}{{3t^{\te\tm}}^{1/3}}$ \\

\noindent ${\cal C}$i)   \> 
$t^{\sr\se\sc} =  - \frac{1}{2\sqrt{2R - R_{\tT\to\ttt\ta\tl}}}\mbox{ln}
\left( 
\frac{2}{t^{\te\tm}} + \frac{1}{\sqrt{2R - R_{\tT\to\ttt\ta\tl}}}
\right) + \mbox{const} $ \>
$t^{\sr\se\sc} =  - \frac{1}{2\sqrt{2R - R_{\tT\to\ttt\ta\tl}}}\mbox{ln}
\left( 
\frac{2}{t^{\te\tm}} + \frac{1}{\sqrt{2R - R_{\tT\to\ttt\ta\tl}}}
\right) + \mbox{const} $ \\

\noindent  ${\cal C}$ii) \>  $t^{\sr\se\sc} =  \mbox{const} + 
\frac{\mbox{\sll\sn}\,t^{\te\tm}}{2\sqrt{2R - R_{\tT\to\ttt\ta\tl}}}$ \>
$t^{\sr\se\sc} =  \mbox{const} + 
\frac{\mbox{ln}t^{\te\tm}}{2\sqrt{2R - R_{\tT\to\ttt\ta\tl}}}$  \\

\noindent  ${\cal C}$iii) \> $t^{\sr\se\sc} =  - \frac{1}{2\sqrt{2R - R_{\tT\to\ttt\ta\tl}}}
\mbox{ln}
\left( 
\frac{2}{t^{\te\tm}} + \frac{1}{\sqrt{2R - R_{\tT\to\ttt\ta\tl}}}
\right) + \mbox{const}$    \>

$t^{\sr\se\sc} =  - \frac{1}{2\sqrt{2R - R_{\tT\to\ttt\ta\tl}}}
\mbox{ln}
\left( 
\frac{2}{t^{\te\tm}} + \frac{1}{\sqrt{2R - R_{\tT\to\ttt\ta\tl}}}
\right) + \mbox{const}$ \\

\noindent ${\cal C}$vi) \>  $t^{\sr\se\sc} = \frac{1}{R_{\tT\to\ttt\ta\tl} - 2R}\mbox{arctan}
\left(
\frac{t^{\te\tm}}{R_{\tT\to\ttt\ta\tl} - 2R}
\right)  
+ \mbox{const}  $                 \>
$t^{\sr\se\sc} = \frac{1}{R_{\tT\to\ttt\ta\tl} - 2R}\mbox{arctan}
\left(
\frac{t^{\te\tm}}{R_{\tT\to\ttt\ta\tl} - 2R}
\right)  
+ \mbox{const}  $                 \\

\noindent ${\cal D}$) \> \> 
$t^{\sr\se\sc} = \mbox{const} - 
\frac{7}{\{2E\}^{26/21}}\{\frac{3}{4}\}^{7/8}\frac{1}{\{t^{\te\tm}\}^{1/7}}$ \\

\noindent ${\cal E}$i) \> Not explicitly evaluable \> Not explicitly evaluable \\

\noindent ${\cal E}$ii)\> Not explicitly evaluable \> Not explicitly evaluable  \\

\end{tabbing}

\noindent
All the above evaluable rectified timefunctions are, additionally, invertible and monotonic. 
As regards interpreting the rectified timefunction, in each case using $t^{\sr\se\sc}$ amounts to 
working on the shape space itself, i.e. using the geometrically natural presentations of Sec 3.1.   
Finally, approximately isotropic GR has an analogue of rectifiation too, amounting to absorption of 
extra factors of the scalefactor $a$ viewed as a function of $t^{\se\sm}$.

\section{Conclusion}

\subsection{This paper's RPM results} 

Euclidean relational particle mechanics (ERPM) is a mechanics in which only relative times, relative 
angles and relative separations are meaningful, and similarity relational particle mechanics (SRPM) 
is a mechanics in which only relative times, relative angles and {\sl ratios} of relative separations 
are meaningful.  
The RPM of N particles in 1-$d$ is N-stop metroland, and that in 2-$d$ is N-a-gonland, of which the 
first two nontrivial cases are triangleland and quadrilateralland.

In this paper, I considered the structure of the configuration space of ERPM (`relational space').  
It is the cone over the configuration space (`shape space') of the corresponding SRPM. 
Thus, following from previous work on the latter \cite{FORD, Kendall}, the N-stop metroland relational 
spaces are C($\mathbb{S}^{\sN - 2}$) and C($\mathbb{RP}^{\sN - 2}$) for plain and oriented shapes 
respectively, with C($\mathbb{CP}^{\sN - 2}$) and C($\mathbb{CP}^{\sN - 2}/\mathbb{Z}_2$) as their 
N-a-gonland counterparts.
The triangleland case's shape space is $\mathbb{CP}^1$, which also admits a further $\mathbb{S}^2$ 
presentation.  
I consider the topological and metric structure of these configuration spaces toward understanding 
classical and quantum ERPM \cite{scaleQM, 08III, MGM, SemiclIII}.    
Despite 4-stop metroland and triangleland both involving (cones over) $\mathbb{S}^2$, they 
are substantially differently realized.  
For example, 4-stop metroland has $\iota = \sqrt{I}$ (the square root of the total moment of inertia) as 
its `radial' variable, while the $\mathbb{S}^2$ presentation of triangleland has, rather, $I$ itself 
in this role. 
Furthermore 4-stop metroland's three ratios are straightforwardly related to the Cartesian directions 
in the $\mathbb{R}^3 = \mC(\mathbb{S}^2)$ relational space, but the corresponding Cartesian directions 
for Triangleland have a rather more complicated meaning (`Dragt coordinates') in terms of the ratio of 
relative Jacobi vector magnitudes and the relative angle between the relative Jacobi vectors that 
conveniently characterize the dynamics of the scalefree triangle.
I extend \cite{AF, +tri}'s geometrical characterization of the above two cases in terms of shape 
quantities to 3-stop metroland and to cases with scale.    
I likewise extend the method of physical interpretation by tessellation of the configuration space, 
which is useful for subsequent reading off of the significance of classical trajectories and QM 
wavefuctions (see also \cite{AF, +tri, scaleQM, 08III, MGM, SemiclIII, Tpaper}).

I have shown that consideration of relational spaces also allows for a direct first-principles 
construction of ERPM as opposed to the indirect formulation in which these theories were first conceived.  
The former involves the natural construction of a mechanics given a (here Riemannian) geometry, along 
the lines of Jacobi and of Synge \cite{Lanczos}.    
These two schemes coincide in the 1- and 2-$d$ cases investigated in the present paper, as reducing 
the latter reproduces the former.  
Moreover, there are differences between these coincident schemes and the relational-absolute split of 
Newtonian Mechanics (although these share much useful kinematics, in particular as regards finding 
useful coordinate systems, e.g. Jacobi coordinates and Dragt-type coordinates).  
Firstly, RPM's have zero total angular momentum, which within Newtonian Mechanics is a simpler and 
qualitatively different case from nonzero total angular momentum.
Nonzero total angular momentum Newtonian Mechanics e.g. gives rise to configuration space monopole 
defects associated with Guichardet-type connections stemming from the rotation group, but these are 
absent in the zero total angular momentum case in analogy with how uncharged particles do not `feel' 
magnetic monoples.  
[I have also established that translational and dilational analogues of the Guichardet connection are 
trivial and so do not contribute any further defect issues relevant to RPM's.]
Secondly, even for the zero total angular momentum case, I point out that there is a difference between 
the relational part of relational-absolute split mechanics and purely relational mechanics at the level 
of the quantum-mechanical operator ordering; the first is that used in the Molecular Physics literature, 
while the second is coincident, or closely related to, operator orderings that are fairly commonly used 
in Quantum Cosmology [Laplacian ordering, conformal ordering and $\xi$-ordering: $D^2 - 
\xi\,\mbox{Ric}(M)$].  
This is interesting from the perspective of the `absolute versus relative motion' debate. 
It reflects that there is an `absolute imprint' difference between the case in which some particles 
constituting a molecule within absolute space (or, implicitly, a much larger universe with local 
conditions being conducive to having a nice stable reference system as is the case e.g. on Earth) and 
the case in which the same particles constituting a whole model universe. 
N.B. that this difference is present even in the simpler and qualitatively distinct case of zero total 
angular momentum (for which, on the one hand, the split of Newtonian Mechanics simplifies, and, on the 
other hand, explicit relational models are actually available for comparison.)

Unlike in minisuperspace models, scalefree RPM's have considerable freedom in the form of the potential 
and scaled RPM's have complete freedom.  
I elect to use up this freedom by choosing potentials which closely parallel those for various 
well-known GR cosmologies; in particular, I choose to emulate the `dominant scale dynamics' part of 
cosmological models by matching choice of potential, and let this naturally induce what the shape part of the 
potential is to be.   
This parallel is via the Mechanics--Cosmology analogy as applied to RPM's, which works out differently 
for each of N-stop metroland and triangleland due to the difference in radial variable between these 
two cases.  
For 1-d RPM's and the $\mathbb{CP}^{\sN - 2}$ presentation of 2-d RPM's, the analogy that   
$-2E$ corresponds to the spatial curvature $k$. 
A combination of Hooke's coefficients from the (upside-down) HO terms, $-2A$, corresponds to the 
cosmological constant term $\Lambda/3$. 
In particular, I identify previous work on (upside down) HO (type) potentials \cite{08I, 08II, AF, +tri, 
MGM, SemiclI} as corresponding to the `Milne in anti de Sitter' oscillating cosmological model and de 
Sitter/inflationary type models in the case of zero rational momentum and with extra wrong-sign 
radiation in the nonzero rational momentum case.  
$-{\cal R}_{\sT\so\st\sa\sll}$ --- minus the total relative rational momentum --- corresponds to a 
cosmological radiation term, but of the `wrong sign'; $2R - {\cal R}_{\sT\so\st\sa\sll}$ is the overall 
right-or-wrong sign radiation term, $2R/\iota^4$ being the surviving term from the conformally-invariant 
potential terms $-R/\iota^2$ in the case of approximate negligibility of shape terms, and corresponding 
to the matter term for radiation, $2GM/a^4$.
If this term is large enough, one has overall a right-sign radiation term.  
$-2K/\iota^3$ -- the surviving term from Newtonian potential terms $K/\iota$ in the case of approximate 
negligibility of shape terms -- corresponds to a dust matter term, $2GM/a^3$.

For the spherical presentation of triangleland, however, while $-{\cal R}_{\sT\so\st\sa\sll}$ and the 
surviving term from $r_{IJ}\mbox{}^{-2}$ potentials play the same roles again, it is $2A$ that now plays 
the role of spatial curvature $k$, and so is conventionally rescaled to be --1, 0 or 1 (which rescaling 
applies implicitly also throughout all the other triangleland--Cosmology analogies). 
The energy $E$ now plays the role of $GM$ for dust, and the surviving term from $r_{IJ}\mbox{}^6$ 
potentials plays the role of the cosmological constant term $\Lambda/3$.  
The problem with $E$ and $A$ only maps to the Newton-Coulomb problem (both attractive and repulsive 
signs can occur); for zero rational momentum this is a dust and curvature universe while for nonzero 
rational momentum, it has also a wrong-sign radiation term.   
Adding enough counterbalancing ${\cal R}$ turns this into a right-sign radiation term.  
Inclusion of radiation terms is desirable as regards modelling simple hot early universes; `wrong signs' 
here do also occur in some more conventionally exotic scenarios involving `dark radiation' \cite{DarkRad}.  
Inclusion of cosmological constant terms allow for RPM's to e.g. mimic simple inflationary universes 
(and simple models of late-universe acceleration).

\subsection{Further work on RPM's}

I present the QM of N-stop metroland in \cite{scaleQM}, and that of triangleland in \cite{08III}; 
these are supported by the present paper's operator ordering and lack of monopole effect results, 
as well as considering some of the present paper's new cosmologically-motivated potentials.  
I also compare Dirac and reduced/relationalspace approaches in \cite{RedDir}.  
If Sec 10.3's applications prove to be fruitful for N-stop metroland and triangleland, I will also 
consider the quadrilateral that unifies the most useful aspects of 4-stop metroland and triangleland; a 
brief kinematical study of quadrilateralland will be presented in \cite{QShape}.  
I also note that C($\mathbb{CP}^{2}$) and C($\mathbb{CP}^{2}/\mathbb{Z}_2$) are rather less trivial 
than C($\mathbb{S}^{2}$) and C($\mathbb{S}^{2}/\mathbb{Z}_2$) making the present paper and its 
extension to these further examples of long-term importance to the RPM program.

\subsection{Problem of Time applications in and of this paper}

For the abovementioned range of potentials and subsequent (approximate) classical solutions for the 
scale part, I consider what the approximate timestandards for the semiclassical approach are.   
The further issue of in what regions of configuration space various conditions for the semiclassical 
approach apply well, I address in \cite{SemiclIII} using the tessellation by the physical interpretation 
method provided in the present paper and in \cite{AF, +tri}.  
The resulting time-dependent Schr\"{o}dinger equation for the light, fast L-part is simplest if one 
passes to the rectified emergent time as provided in the present paper.  
Solving the resulting time-dependent Schr\"{o}dinger equation (with further approximations being allowable 
as rectified emergent time 
dependent perturbations) I consider in \cite{SemiclIII} (with a small foretaste in \cite{MGM}).  
Solving this coupled to a less approximate H equation (small QM expectation type 
perturbations about a Hamilton--Jacobi equation), which is perhaps to be seen as \cite{SemiclI, 
SemiclIII} (an extension of) the well-known Hartree--Fock approach to Atomic and Molecular Physics, 
is a more complicated such scheme that allows for backreaction of the L subsystem on the H one and the  
approximate timestandard that it provides.    
Throughout the above, checks against ulteriorly exactly soluble cases documented in \cite{08II, AF, 
+tri, scaleQM, 08III} are useful checks of the semiclassical approach's assumptions and approximations; 
moreover, such ulterior exact solvability is seldom available in minisuperspace.  
Also, the relative simpleness of the L-dynamics of shape for RPM's is useful in investigating various 
of the above-outlined features. 
Thus, RPM's may be viewed as valuable toy models of midisuperspace Quantum Cosmology models tied to the 
origin of structure formation in the universe (e.g. the Halliwell--Hawking model toward 
Quantum Cosmology seeding galaxy formation and CMB inhomogeneities).
[Thus RPM's are valuable conceptually and to test whether we should or should not be {\sl qualitatively} 
confident in the assumptions and approximations made in such schemes.]

I also consider dilational Euler hidden time in \cite{IT}, and timeless approaches in \cite{scaleQM} 
(na\"{\i}ve Schr\"{o}dinger interpretation) and \cite{NOD, NOI} (records theory), and aim to consider 
whether records theory, histories theory and the semiclassical approaches to the Problem of Time can 
be combined.   
The present paper's `52 analogies between RPM's and GR-as-geometrodynamics' Section contains a number of 
further questions about Problem of Time and foundations of Quantum Cosmology including the problem of 
observables, the issue of uniform states, the origin of the arrow of time and robustness issues as 
regards neglecting some of a model universe's modes.  


\mbox{ }  

\noindent{\bf Acknowledgements}  

\mbox{ }

\noindent
I thank my wife Claire and my friends Alicia, Amelia, Coryan and Ed for cheering me up during the 
period in which this work was carried out.    
I thank Professors Jeremy Butterfield, Gary Gibbons, Malcolm MacCallum, Don Page and Reza Tavakol, 
and Dr Alexei Grinbaum, for help with my career/having me as an academic visitor.   
I also thank Miss Anne Franzen, Dr Jonathan Oppenheim, Mr Simeon Bird, Mr Sean Gryb, Mr Henrique Gomes, 
and Professors Gary Gibbons, Harvey Brown, Julian Barbour and Jeremy Butterfield for comments and 
discussions.

\mbox{ }

\noindent{\bf\Large Appendix `Guichardet connection' for various transformation groups}

\mbox{ }  

\noindent{\bf\large 1 The Guichardet connection for rotations}  

\mbox{ }

\noindent 
Working in mass-weighted Jacobi coordinates, 
\beq
\underline{\ttL} = 
\sum\mbox{}_{\mbox{}_{\mbox{\scriptsize $i = 1$}}}^{\sn}\underline{\iota}^i \cr \underline{\pi}_i = 
\sum\mbox{}_{\mbox{}_{\mbox{\scriptsize $i = 1$}}}^{\sn}\underline{\iota}^i \cr 
\{\underline{\dot{\iota}}^i + \underline{\dot{\mB}} \cr \underline{\iota}^i\} = 
\sum\mbox{}_{\mbox{}_{\mbox{\scriptsize $i = 1$}}}^{\sn}\underline{\iota}^i \cr 
\big\{
\dot{\cal Q}^{\Delta}{\pa\underline{\iota}^i}/{\pa {\cal Q}^{\Delta}}
 + \underline{\dot{\mB}} \cr \underline{\iota}^i
\big\} 
\eeq
(for $\underline{\pi}_i$ the momentum conjugate to $\underline{\iota}^i$), then let
\beq
\underline{\ttL} = \underline{\underline{I}}\,\{\underline{\dot{\mB}} + 
\underline{A}_{\Delta}\dot{\cal Q}^{\Delta}\}
\eeq
for $\underline{\underline{I}}$ the inertia tensor.  
The last term in this defines the Guichardet-type \cite{Guichardet} gauge potential $\underline{A}_{\Delta} = 
\underline{\underline{I}}^{-1}\underline{a}_{\Delta}$ for $\underline{a}_{\Delta} = 
\sum_{i = 1}^{\sn}\underline{\iota}^{i} \cr {\pa \underline{\iota}^{i}}/{\pa {\cal Q}^{\Delta}}$.  
For vanishing angular momentum, ${\dot{\underline{\mB}}} = - \underline{A}_{\Delta}\dot{\cal Q}^{\Delta}$ i.e. the 
mapping between change of shape (taken to include scale in this useage) and corresponding 
infinitesimal rotation.  
In 2-$d$ and for unoriented scaled triangleland,

\noindent using ${\cal Q}^{\Delta}$ =($\iota^1$, 
$\iota^2$, $\Phi$) coordinates (closely related to parabolic coordinates \cite{08I,08III}) and the `xxy gauge' 
\cite{LR} in which $\underline{\iota}^1 = \iota^1(1, 0)$ and 
$\underline{\iota}^2 = \iota^2(\mbox{cos}\,\Phi, \mbox{sin}\,\Phi)$, the nonzero 
component of $A_{\Delta}$ is 
\beq
{A}_{\Phi} = \mbox{Tall} \mbox{ } , 
\eeq
which, in passing to Dragt coordinates, gives 
\beq
\underline{A}_{\Delta}\d {\cal Q}^{\Delta} = \mbox{$\frac{1}{2}$}\{
     { \mbox{Dra}^1\d \mbox{Dra}^2 - \mbox{Dra}^2\d \mbox{Dra}^1 }\}/
     {I\{I - \mbox{Dra}^3\}    } \mbox{ } .  
\eeq
This is in direct correspondence with Wu--Yang's \cite{YW} $A^{\sS}_{\alpha}$ for the Dirac monopole if one passes 
from $x^{\alpha}$ to $\mbox{Dra}^{\Delta}$, sets the monopole strength $g$ to be 1/2 and uses the 
clustering in question's M in place of S in its role of defining a chart that does not cover this gauge's 
manifestation of the Dirac string (which runs along the opposite oriented clustering's D-axis).  
Likewise, if one inverts the roles of $\underline{\iota}^1$ and $\underline{\iota}^2$, the 
nonzero component of $A_{\Delta}$ in the resulting chart and gauge is 
\beq
{A}_{\Phi} = \mbox{Flat} \mbox{ } , 
\eeq
which, in passing to Dragt coordinates, gives 
\beq
{A}_{\Delta}\d {\cal Q}^{\Delta} = \mbox{$\frac{1}{2}$}
\{   {\mbox{Dra}^1\d \mbox{Dra}^2 - \mbox{Dra}^2\d \mbox{Dra}^1}\}/  
     {I\{I + \mbox{Dra}^3\}    } \mbox{ } .  
\eeq
This is in direct correspondence with Wu--Yang's $A^{\sN}_{\alpha}$ for the Dirac monopole if one passes 
from $x^{\alpha}$ to $\mbox{Dra}^{\Delta}$, sets the monpole strength $g$ to be 1/2 and uses the 
clustering in question's D in place of N in its role of defining a chart that excludes this gauge's 
manifestation of the Dirac string (which now runs along the opposite orientation's clustering's M-axis).
Then this D-chart and M-chart provide a full stringless description of this relational space 
monopole, just as the N-chart and S-chart do for the usual Dirac monopole in space.  
Given the precise nature of the correspondence between these, it is clear that the field strength is 
\beq
F_{\Gamma\Delta} =  \epsilon_{\Lambda\Gamma\Delta}{\mbox{Dra}^{\Lambda}}/{I^3} \mbox{ } .  
\label{nontri}
\eeq

For oriented scaled triangleland, these workings still hold except that the $\mbox{Dra}^3 = 
\mbox{TetraArea} < 0$ half-plane has ceased to be part of the configuration space and that other 
charts and gauges which position the Dirac string elsewhere are now more convenient; see e.g. 
\cite{Iwai, LR}.  
In this case one has an {\sl Iwai monopole} on $\mathbb{R}^3_+$.   

\mbox{ }  

\noindent{\bf\large 2 Translations and dilations give but trivial analogues}  

\mbox{ }

\noindent The below results hold for all particle numbers and spatial dimensions.

In the case of translations, the coordinates are mass-weighted particle positions, 
$\underline{X}^I = \sqrt{m_I}\underline{q}^{I}$ rather than mass-weighted Jacobi coordinates.  
\beq
\ttP_{\mu} = \sum\mbox{}_{\mbox{}_{\mbox{\scriptsize $I = 1$}}}^{\sN}\underline{\pi}_I = 
\sum\mbox{}_{\mbox{}_{\mbox{\scriptsize $I = 1$}}}^{\sN}m_I\{\underline{q}_{I} + \underline{\dot{\mA}}\} = 
M \sum\mbox{}_{\mbox{}_{\mbox{\scriptsize $I = 1$}}}^{\sN} \big\{ \underline{\dot{\mA}} +  
M^{-1}\sum\mbox{}_{\mbox{}_{\mbox{\scriptsize $I = 1$}}}^{\sN} 
\sqrt{m_I}\dot{\cal Q}^{\Delta}{\pa\underline{X}^I}/{\pa {\cal Q}^{\Delta}} \big\} 
\eeq
Here, the total mass is the analogue of inertia tensor, so 
$\underline{A}^{\st\sr\sa\sn\sss}_{\Delta} = \underline{a}^{\st\sr\sa\sn\sss}_{\Delta}$  
$ = \pa_{\Delta}\{  \sum_{I = 1}^{\sN}\sqrt{m_I}\underline{X}^{I}/M \}$.  
As this is of gradient form $\pa_{\Delta}\zeta$, the corresponding translational field strength 
$F_{\Gamma\Delta} = 2\pa_{[\Gamma}\pa_{\Delta]}\zeta = 0$ by symmetry--antisymmetry.  
Thus this connection is flat/geometrically trivial.

In the case of dilations, taking the coordinates to be mass-weighted Jacobi coordinates, 
\beq
\ttD = \sum\mbox{}_{\mbox{}_{\mbox{\scriptsize $i = 1$}}}^{\sn}\underline{\iota}_i \cdot \underline{\pi}_i = 
\sum\mbox{}_{\mbox{}_{\mbox{\scriptsize $i = 1$}}}^{\sn}\underline{\iota}_i\cdot\{\underline{\iota}_i + 
{\dot{\mC}}\underline{\iota}_i\} = 
\sum\mbox{}_{\mbox{}_{\mbox{\scriptsize $i = 1$}}}^{\sn}\underline{\iota}_i \cdot 
\big\{ \dot{\cal Q}^{\Delta}{\pa\underline{\iota}_i}/{\pa {\cal Q}^{\Delta}} + 
{\dot{\mC}}\underline{\iota}_i \big\} \mbox{ } , 
\eeq
so
\beq
\ttD =  I\{\dot{\mC} +  {A}_{\Delta}\dot{\cal Q}^{\Delta}\} \mbox{ } .  
\eeq
Here the scalar moment of inertia $I = \sum_{i = 1}^{\sn}\{\iota^i\}^2$ is the analogue of the inertia tensor, 
and ${A}^{\sd\si\sll}_{\Delta} = I^{-1}a^{\sd\si\sll}_{\Delta}$, $a^{\sd\si\sll}_{\Delta} = 
\sum_{i = 1}^{\sn}\underline{\iota}^i\cdot{\pa \underline{\iota}^i}/{\pa {\cal Q}^{\Delta}}$.  
Then for vanishing dilation, $\dot{\mC} = - A^{\sd\si\sll}_{\Delta}\dot{\cal Q}^{\Delta}$ so it is a 
mapping between change of shape and corresponding infinitesimal size change.  
But again this can be cast in gradient form: $A^{\sd\si\sll}_{\Delta} = 
\pa_{\Delta}\{ \mbox{ln}(\iota)\}$, so the corresponding field strength is also zero and so this 
connection is also flat/geometrically trivial.  

\noindent Finally, composition of translational, rotational and dilational corrections is additive, so 
outcomes for each of these things do not affect each other.   
(To consider combinations involving the translations, 
note that the above presentations for mass-weighted relative Jacobi coordinates for rotations and  
dilations continue to hold identically under $i$ to $I$, n to N = n + 1 and $\iota^i$ to $X^{I}$ 
-- trivial position to relative Jacobi coordinates map, see e.g. \cite{06I, 06II}.)


\end{document}